\RequirePackage{fix-cm}
\RequirePackage{amsmath, bbm}
\documentclass[twocolumn,epjc3]{svjour3}
\pdfoutput=1
\smartqed  % flush right qed marks, e.g. at end of proof
\usepackage{tabularx} % in the preamble
\usepackage{graphicx}
\usepackage{axodraw2}
\usepackage{amscd}
\usepackage{slashed}
\usepackage{multirow}
\usepackage{float}
\usepackage[caption=false]{subfig}

\journalname{Eur. Phys. J. C}
\begin{document}

\title{Sensitivity of Future
  Hadron Colliders to Leptoquark Pair Production in the Di-Muon Di-Jets Channel%\thanksref{t1}
}


\author{B. C. Allanach\thanksref{addr1}
        \and
        Tyler Corbett\thanksref{addr2}
	\and
	Maeve Madigan\thanksref{e1,addr1}
}

\thankstext{e1}{e-mail: mum20@cam.ac.uk}


\institute{DAMTP, University of Cambridge, Wilberforce Road, Cambridge, CB3 0WA, United Kingdom \label{addr1}
           \and
           The Niels Bohr International Academy, Blegdamsvej 17, University of Copenhagen, DK-2100 Copenhagen, Denmark \label{addr2}
}

\date{Received: date / Accepted: date}

\maketitle

\begin{abstract}
We estimate the future sensitivity of the\\
high luminosity (HL-) and high energy (HE-) modes of the Large Hadron Collider
(LHC) and of a 100 TeV future circular
collider (FCC-hh) to
leptoquark (LQ) pair production in the
muon-plus-jet decay mode of each LQ\@.
Such LQs are motivated by the fact that they
provide an explanation
for
the neutral current $B-$anomalies. For each future collider, Standard Model (SM) backgrounds
and detector effects are simulated. From these, sensitivities of each collider are found.
Our measures of sensitivity are based upon a Run II ATLAS search, which we also use for validation.
We illustrate with a narrow scalar (`$S_3$') LQ and
find that, in our channel, the HL-LHC
has exclusion sensitivity to LQ masses up to
1.8 TeV,
the HE-LHC up to 4.8 TeV and the
FCC-hh up to 13.5 TeV.
\end{abstract}
\section{Introduction}
\label{sec:intro}

There has been much attention recently on various measurements in rare
$B$-meson decays, since several of them are in apparent disagreement with
Standard Model\\ (SM) predictions. The measurements that we focus on here
involve processes with muon pairs, bottom quarks and strange
quarks. Most of the disagreements are at the $2-3 \sigma$ level and so do not
warrant particular consternation of and by themselves. However, if one takes
them collectively,
it seems as if a pattern may be emerging. Of particular interest are the
ratios of
branching ratios
\begin{align}
R_{K^{(\ast)}} \equiv \frac{BR(B \rightarrow K^{(\ast)} \mu^+ \mu^-)}
{BR(B\rightarrow K^{(\ast)} e^+ e^-)},  
  \end{align}
both predicted by the SM to be 1.00 with high precision
in the di-lepton invariant mass squared bin
$m_{ll}^2 \in [1.1,\ 6]$ GeV$^2$.
LHCb measurements~\cite{Aaij:2017vbb,CERN-EP-2019-043} imply
$R_K=0.846^{+0.060}_{-0.054}{}^{+0.016}_{-0.014}$ and
$R_{K^{\ast}}=0.69^{+0.11}_{-0.07}\pm0.05$ in this bin.
Other observables disagree with their SM prediction, despite larger
theoretical uncertainties.
For example, the branching ratio of $B_s \rightarrow \mu^+ \mu^-$~\cite{Aaboud:2018mst,Chatrchyan:2013bka,CMS:2014xfa,Aaij:2017vad} is also measured to be lower than the SM
prediction. Discrepancies with SM
predictions~\cite{Khachatryan:2015isa,Bobeth:2017vxj} include some of the angular
distributions in $B \rightarrow K^{(\ast)} \mu^+ \mu^-$
decays~\cite{Aaij:2013qta,Aaij:2015oid,ATLAS-CONF-2017-023,CMS-PAS-BPH-15-008}.
Collectively, we refer to the discrepancies between these measurements and SM
predictions as the neutral current $B-$anomalies (NCBAs)\footnote{In particular, here
we shall not motivate new particles by the charged current anomalies in
$B \rightarrow D^{(\ast)} \tau \nu$ decays.}.

We shall take the hypothesis that the NCBAs are harbingers of new
particles with flavour dependent
interactions. We suppose that the new particles are much heavier than $B$ mesons, such that
they will not have
necessarily already been produced and discovered in previous high energy colliders. In SM effective
theory, global fits find that the two most relevant beyond the SM (BSM) Lagrangian operators for
describing such new particles are~\cite{Alguero:2019ptt,Alok:2019ufo,Ciuchini:2019usw,Aebischer:2019mlg,Kowalska:2019ley,Arbey:2019duh}
\begin{align}  \label{bsmOp}
	{\mathcal L}_{WET}&= \frac{1}{(36\text{~TeV})^2} [
C_L (\overline{s_L} \gamma_\rho b_L )
        ( \overline{\mu_L} \gamma^\rho \mu_L )\\ 
&+ C_R (\overline{s_L} \gamma_\rho b_L )
( \overline{\mu_R} \gamma^\rho \mu_R )
]+h.c.  \nonumber
\end{align}
The $(1/36)^2$ TeV normalisation makes the $C_{L,R}$ dimensionless, and for
$C_{L,R} \sim {\mathcal O}(1)$ they can
fit the NCBA data.
Bearing Eq.~\ref{bsmOp} in mind, the only tree-level solutions explaining
these deviations are those of $Z'$s and leptoquarks. Pioneering projections for the direct
discovery of such new particles at the LHC and future colliders, in a
simplified framework, were made in Ref.~\cite{Allanach:2017bta,Hiller:2018wbv}.
More accurate estimates involving simulations of collisions and detector
response were made
in Ref.~\cite{Allanach:2019mfl,Allanach:2018odd} for the $Z^\prime$ case.
Heavy LQs  with
couplings to $b$ quarks, $s$ quarks and muons predict the BSM operators in
Eq.~\ref{bsmOp}~\cite{Deppisch:2016qqd,Capdevila:2017bsm,Hiller:2017bzc,DAmico:2017mtc,Alda:2018mfy,Kumar:2018kmr,Alvarez:2018gxs,DaRold:2019fiw}. Here
we shall consider the easier example of
a scalar LQ. Vector leptoquarks can also successfully explain the NCBAs, but to obtain sensible high
energy behaviour, require a full ultraviolet model to be devised, introducing
further model dependence.

A complex $SU(2)_L$ triplet scalar
$S_3$, with quantum numbers $(\bar 3,3,\frac{1}{3})$ under the standard model gauge group $SU(3)_c\times
SU(2)_L\times U(1)_Y$, can produce the BSM operator with coefficient $C_{L}$ in Eq.~\ref{bsmOp}. In
fact, it is the only single scalar LQ progenitor.
There are two alternative vector boson LQ
progenitors: $U_1$  and $U_3$ with quantum numbers $(\bar
3,1,\frac{2}{3})$ and $(3,3,\frac{2}{3})$, respectively.
Many of our results (for example, bounds on production cross section times
branching ratio) shall apply equally to other LQs such as $U_1$ and $U_3$. However,
we shall illustrate  some other estimates
that depend on calculation of signal
cross section and branching ratio (for example, sensitivity to $m_{\text LQ}$)
solely for the $S_3$ case, vector LQ
simulations being beyond the scope of the present paper, since they add model dependence
concomitant with the necessity of providing a more complete ultraviolet
model\footnote{Vector leptoquarks' couplings to gluons depend on a free parameter in contrast to scalar leptoquarks whose interactions are fixed by the QCD coupling.}.
\begin{figure}
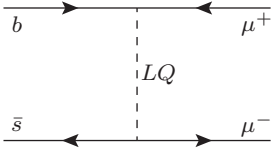

\begin{center}
\begin{axopicture}(100,50)(0,0)
\Text(58,25)[c]{$LQ$}
\Text(5,6)[c]{$\bar s$}
\Text(5,44)[c]{$b$}
\Text(95,44)[c]{$\mu^+$}
\Text(95,6)[c]{$\mu^-$}
\Line[arrow](50,0)(0,0)
\Line[arrow](50,0)(100,0)
\Line[arrow](0,50)(50,50)
\Line[arrow](100,50)(50,50)
\Line[dash](50,0)(50,50)
\end{axopicture}
\end{center}
\caption{Feynman diagram showing LQ mediation of an effective operator
contributing to NCBAs. \label{fig:LQmed}}
\end{figure}
The Yukawa couplings of the LQ to the $i^{th}$-family SM quark (${Q'}_i$) and
lepton (${L'}_i$) $SU(2)_L$ doublets are given by (in the primed weak eigenbasis) \cite{Dorsner:2016wpm,deMedeirosVarzielas:2019lgb}:
\begin{align}\label{S3yukawas}
	\mathcal L_\text{Yukawa}&= (Y_L)_{ij}\overline{Q^{C\prime}}_{i,a} \epsilon_{ab}\tau^k_{bc}{L'}_{j,c} S_3^k\\ &+  (Y_Q)_{ij}\overline{Q^{C\prime}_{i,a}} \epsilon_{ab}\tau^k_{bc} {Q'}_{j,c} (S_3^k)^\dagger+{h.c.},\nonumber
\end{align}
where we have suppressed QCD gauge indices, $i, j \in \{1,2,3\}$ are family
indices (repeated indices have an implicit summation
convention), $a, b, c \in \{1,2\}$ are fundamental $SU(2)_L$ indices, $k \in
\{1,2,3\}$ is an adjoint $SU(2)_L$ index, the superscript $C$ denotes a charge conjugated fermion, $\epsilon_{ab}$ is the Levi Civita symbol, the $\tau^k_{ab}$ are
the Pauli matrices and $Y_L$ and $Y_Q$ are 3 by 3 matrices
of complex dimensionless Yukawa couplings.

In order to avoid proton instability we assume that baryon number is conserved,
setting $(Y_Q)_{ij}$ to zero in consequence.
After electroweak symmetry breaking, $S_3$ becomes $(S^{-2/3},\ S^{+1/3},\
S^{+4/3})$ where the superscript denotes electric charge. The left-handed
quarks and leptons mix according to
\begin{align}
{{\bf P}'}^T = V_P {\bf P}^T,
\end{align}
where\footnote{Here, the transpose denotes a column vector.} $P \in \{u_L,\ d_L,\ e_L,\ \nu_L\}$, bold face denotes a 3-vector in
family space and
the unprimed basis is the mass eigenbasis. $V_P$ are then unitary
dimensionless 3 by 3 matrices, being experimentally constrained via the
Cabbibo-Kobayashi-Maskawa
combination\\$V_{CKM}=V_{u_L}^\dag V_{d_L}$ and the
Pontecorve-Maki-Nakagawa-Sakata combination $U_{PMNS}=V_{\nu_L}^\dag
V_{e_L}$. In the mass eigenbasis,
\begin{align}  \label{unPrimedS3}
	\mathcal L_\text{Yukawa}&=
-\sqrt{2}\overline{{\bf  d}_L^C} Y_{de}  {\bf e}_LS^{+4/3}
-\overline{{\bf u}_L^C} Y_{ue}  {\bf  e}_LS^{+1/3}\\ 
	&-\overline{{\bf d}_L^C}Y_{d\nu}{\bf \nu}_LS^{+1/3}
+\sqrt{2}{\bf u}_L^C Y_{u\nu}  {\bf \nu}_LS^{-2/3} 
+ h.c.,  \nonumber
\end{align}
where $Y_{de} = V_{d_L}^T Y_L V_{e_L}$, $Y_{ue}= V_{u_L}^T Y_L
V_{e_L}$,\\ $Y_{d\nu}=V_{d_L}^TY_LV_{\nu_L}$, and $Y_{u\nu} =V_{u_L}^T Y_L V_{\nu_L}$.
In order to describe the NCBAs, we require $(Y_{de})_{32} \neq 0$ and
$(Y_{de})_{22} \neq 0$.
Then, Fig.~\ref{fig:LQmed}
demonstrates how $S_3$ contributes to the NCBAs via tree-level
exchange.
For LQ masses much larger than $B$ meson masses ($m_{\text LQ} \gg m_B$), the
effective field theory matches
\begin{align}
C_L = (Y_{de})_{32} (Y_{de}^\ast)_{22} \left(\frac{\text{36 TeV}}{m_{\text LQ}}\right)^2. \label{eftMatch}
\end{align}
$C_L=-1.06 \pm 0.16$ fits combined NCBA data~\cite{Aebischer:2019mlg}. We
shall typically use the central value from the fit in order to fix
$(Y_{de})_{32} (Y_{de}^\ast)_{22}$ for a given value of $m_{\text LQ}$. Although in
general $Y_{de}$ are complex, we shall here take real values for simplicity
and because we are not considering $CP$-violating observables.

\begin{figure}
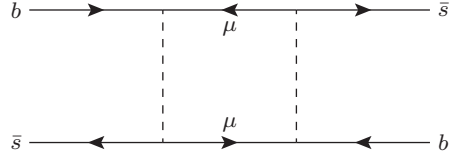

\unitlength=1pt
\begin{center}
\begin{axopicture}(155,50)(-5,0)
\Line[arrow](50,0)(0,0)
\Line[arrow](50,0)(100,0)
\Line[arrow](150,0)(100,0)
\Line[dash](50,50)(50,0)
\Line[dash](100,50)(100,0)
\Line[arrow](0,50)(50,50)
\Line[arrow](100,50)(50,50)
\Line[arrow](100,50)(150,50)
\Text(-5,0)[c]{$\bar s$}
\Text(-5,50)[c]{$b$}
\Text(75,43)[c]{$\mu$}
\Text(75,7)[c]{$\mu$}
\Text(155,0)[c]{$b$}
\Text(155,50)[c]{$\bar s$}
\end{axopicture}
\end{center}
\caption{Leading order Feynman diagram of LQ contribution to $B_s-\bar B_s$ mixing. \label{fig:bsbar}}
\end{figure}
Our LQs contribute to $B_s-{\bar B}_s$ mixing at one-loop order, as shown in
Fig.~\ref{fig:bsbar}.
A recent determination of the SM prediction for $B_s-{\bar B}_s$
mixing is broadly in agreement with the experimental measurement and so upper
bounds can be placed on the LQ contribution. However, a recent
determination~\cite{King:2019lal}
finds that this is not very constraining for $S_3$ leptoquarks that fit the
NCBAs. Perturbative unitarity provides the stronger constraint that $m_{\text LQ}<68$~TeV.

\begin{figure}
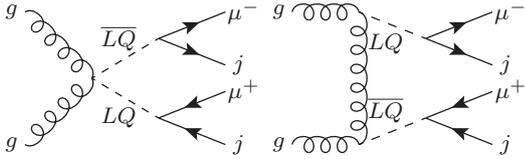

\unitlength=1pt
\begin{center}
\begin{axopicture}(180,50)(-5,0)
% 4 point diagram
\Gluon(0,50)(25,25){3}{4}
\Line[dash](25,25)(50,40)
\Line[dash](25,25)(50,10)
\Line[arrow](75,0)(50,10)
\Line[arrow](75,20)(50,10)
\Line[arrow](50,40)(75,50)
\Line[arrow](50,40)(75,30)
\Gluon(0,0)(25,25){-3}{4}
\Text(-5,50)[c]{$g$}
\Text(-5,0)[c]{$g$}
\Text(82,50)[c]{$\mu^-$}
\Text(80,0)[c]{$j$}
\Text(82,20)[c]{$\mu^+$}
\Text(80,30)[c]{$j$}
\Text(35,40)[c]{$\overline{LQ}$}
\Text(35,10)[c]{$LQ$}

% 3 point diagram
\Gluon(100,0)(125,0){3}{3}
\Gluon(100,50)(125,50){3}{3}
\Gluon(125,50)(125,0){3}{6}
\Line[dash](125,50)(150,40)
\Line[dash](125,0)(150,10)
\Line[arrow](175,0)(150,10)
\Line[arrow](175,20)(150,10)
\Line[arrow](150,40)(175,50)
\Line[arrow](150,40)(175,30)
\Text(182,50)[c]{$\mu^-$}
\Text(180,0)[c]{$j$}
\Text(182,20)[c]{$\mu^+$}
\Text(180,30)[c]{$j$}
\Text(95,50)[c]{$g$}
\Text(95,0)[c]{$g$}
\Text(135,38)[c]{$LQ$}
\Text(135,13)[c]{$\overline{LQ}$}
\end{axopicture}
\end{center}
\caption{Example Feynman diagrams of LQ pair production at a hadron
  collider followed by
  subsequent
  decay of each LQ into $\mu j$. \label{fig:diLQ}}
\end{figure}

At hadron colliders, the dominant mechanism for pair
production of the LQ is by gluon fusion, as shown in
Fig.~\ref{fig:diLQ}. Since by definition, a LQ couples to a quark and a
lepton, it must carry colour to preserve $SU(3)$ and therefore,
by $SU(3)$ gauge symmetry, must
couple to gluons via the QCD coupling constant.
Run II of 13 TeV running at the LHC has produced some 139 fb$^{-1}$ of
integrated luminosity each for ATLAS and CMS\@. A search for
NCBA-solving LQs using all of this data is eagerly awaited, having not
appeared yet.
The next phase of LHC running will be in the 14 TeV
high-luminosity phase (HL-LHC), with a design integrated luminosity of 3
ab$^{-1}$. This phase will provide much new
information on the NCBAs~\cite{Cerri:2018ypt} concurrently with Belle
II~\cite{Albrecht:2017odf,Kou:2018nap}. At the same time, direct searches for
new particles~\cite{CidVidal:2018eel} may include $S_3$. One potential future
LHC upgrade would be to insert 16 Tesla magnets into the current LHC ring,
resulting in the high energy LHC (HE-LHC), running at a nominal energy of 27~TeV with a design integrated luminosity of 15
ab$^{-1}$~\cite{Abada:2019ono}. Ultimately, such magnets could be placed
within a much larger tunnel, resulting in the Future Circular Collider, which
could collide protons (FCC-hh) at a centre of mass energy $\sqrt{s}=100$~TeV with a design luminosity of 20~ab$^{-1}$~\cite{Abada:2019lih,Benedikt:2018csr}.
We arrive at the question central to this paper, which is:
\begin{quote}
{\em
For LQs which fit the NCBAs, what is the LQ mass sensitivity of future hadron
colliders?}
\end{quote}
We wish to estimate the
sensitivity for the Run II, HL-LHC, HE-LHC and FCC-hh options.  Table~\ref{table:colliders} summarises the centre of mass energies and integrated luminosities that will be used for each collider in our estimates.  We hope that this will
help inform the European Strategy for particle physics, which is currently
deliberating on various scientific priorities.
\begin{table}
\begin{center}
\begin{tabular}{|c|c|c|}
 \hline
& $\sqrt{s}$ [TeV] & $\mathcal{L} $ [ab$^{-1}$]   \\
 \hline \hline
LHC &  $13$ & $0.14$  \\
HL-LHC   & $14$ &$3$  \\
HE-LHC &  $27$ & $15$  \\
FCC-hh   & $100$ &$20$  \\
 \hline
\end{tabular}
\caption{Design centre of mass energies and integrated luminosities of the LHC Run II and future hadron colliders.} \label{table:colliders}
\end{center}
\end{table}

A
previous estimate of future collider sensitivity to $S_3$ LQs consistent with
the NCBAs was made in Ref. ~\cite{Allanach:2017bta}, which projected current sensitivity to
higher centre of mass energies and luminosities. However, the sensitivity
estimate had two highly dubious approximations. The first
 was that experimental
efficiency and acceptance did not change with centre of mass energy. In fact,
at large $m_{\text LQ}$ and at high energies (particularly at FCC-hh), the decay products
from LQs will be highly boosted. This has two effects: the muons will be
pushed closer to the jets, meaning that more of them will fail isolation
criteria. Also, at higher energies, the muon momentum resolution is likely to be very
poor, since such hard muons can only be bent to a limited extent by the
magnets. This will also affect the signal efficiency from peak broadening. The second dubious
approximation was that the LQs are produced exactly at threshold. This is
likely to introduce large uncertainties. We shall rectify these
approximations in our paper by performing a fast simulation of the signal and
backgrounds, as well as including
detector response. The first of these approximations has already been found to have
non-trivial effects upon the predicted future hadron collider sensitivity of
$Z^\prime$ explanations of the
NCBAs~\cite{Allanach:2018odd,Jamin:2019mqx}. The estimate in this paper should
be much more accurate than the previous pioneering determination in Ref.~\cite{Allanach:2017bta}.

Searches for LQ pair production with subsequent decays of each into a muon and a
jet have already been performed at the 13 TeV LHC\@.
The ATLAS Collaboration  set a 95$\%$ confidence level lower limit on $m_{\text LQ}$
of 1.05 TeV from 3.2 fb$^{-1}$ of $pp$ collisions~\cite{Aaboud:2016qeg}. This is a simple cut-based
analysis, which we adopt for estimating future hadron collider
sensitivity. More recent experimental analyses were made more sophisticated in
order to squeeze more sensitivity out of them.
The CMS Collaboration maximise their sensitivity using a multi-dimensional optimisation of the final selection for each $m_{\text{LQ}}$
in 36 fb$^{-1}$ of delivered beam at the LHC~\cite{Sirunyan:2018ryt}, finding
a 95$\%$ CL lower bound of $m_{\text LQ}> 1.28$~TeV. The ATLAS collaboration has also
performed a search in 36 fb$^{-1}$ of 13 TeV $pp$ collisions for LQs decaying
to muons and jets. They utilise differential cross-section measurements and
boosted decision trees to
obtain a lower bound of $m_{\text LQ}>1.23$~TeV.
However, such a level of
sophistication is unnecessary for our purposes, where the uncertainties
involved in estimating future collider sensitivities (for example because we
do not yet know the experimental design) are much larger than the gain
in sensitivity. Thus, following the much simpler methodology in Ref.~\cite{Aaboud:2016qeg}
is sufficient for our purposes.  The NCBAs predict that there should be couplings between $S_3$ and $\bar b$, $\mu$ from the first term in Eq.~\ref{unPrimedS3}. Thus we expect a decay channel $S_3 \rightarrow \bar b \mu$ to be open.
In the experimental analysis we choose, the bottom quark remains untagged and is counted merely as a light jet.
We note that the second term in Eq.~\ref{unPrimedS3} may simultaneously predict $S_3$ decays to top quarks and muons. This mode is more complicated than the one we choose for analysis, and we leave it for future work. For a discussion of potential analysis strategies for this decay mode, see Ref.~\cite{Chandak:2019iwj}.

Collider sensitivity to LQ pair production is limited by SM background
rates. Therefore the estimation of such background rates are of vital
importance to the estimate of the sensitivity to LQ pair production.
The paper proceeds as follows:
in \S\ref{sec:back} we describe the SM
backgrounds, how they are simulated.
We validate our estimate of the backgrounds and resulting LQ limits  against
the ATLAS determination for $\sqrt{s}=13$~TeV.
Then we estimate the backgrounds at future hadron colliders.
Next in
\S\ref{sec:sens} we present the sensitivity estimates for future hadron
colliders,
before summarising in
\S\ref{sec:conc}.

\section{Standard Model Backgrounds}
\label{sec:back}
Consider the pair production of LQs and their decay to a $\mu \mu j j$ final state.  Following previous searches for leptoquark pair production, we define the parameter $m_\text{min}(\mu, j)$ from the kinematics of these four final state particles by finding the configuration of muon-jet pairings which minimises the difference in invariant masses $|m(\mu_{1},j_{1}) - m(\mu_{2},j_{2})|$ and choosing $m_\text{min}(\mu,j) =\\ \text{min}[m(\mu_{1},j_{1}), m(\mu_{2},j_{2})]$, where $j_1$ and $j_2$ are the hardest two jets in an event. In an on-shell LQ pair production event this parameter will approximate the LQ mass $m_\text{LQ}$.

To estimate the sensitivity of future colliders to LQ pair production in the $\mu \mu j j$ channel we simulate the distribution of the SM background in $m_\text{min}(\mu,j)$.  We select events containing \textit{exactly two} muons with no charge requirement and \textit{at least two} jets with no flavour requirement. Our background simulations for 13 TeV are validated against the results presented in \cite{Aaboud:2016qeg}.  More details can be found in \S\ref{sec:valid}. We place limits on $\sigma \times \text{BR}$ and determine the maximum LQ mass $m_\text{LQ}$ which could be excluded at $95 \%$ CL by each collider, assuming the observed data is consistent with the SM background.  Alternatively, assuming a LQ exists at mass $m_\text{LQ}$, we estimate the discovery potential by finding the integrated luminosity required for a $5 \sigma$ significance.

\subsection{Methodology} \label{sec:methods}
\begin{figure}
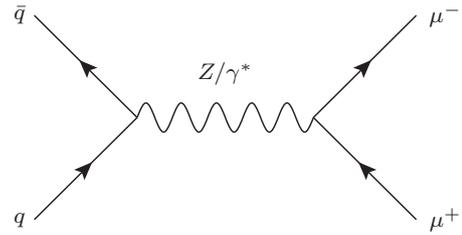
 
\centering
\begin{axopicture}(130,90)
	\SetScale{1.1}
        \Line[arrow](0,0)(35,35)
        \Line[arrow](35,35)(0,70)
        \Photon(35,35)(95,35){5}{5}
        \Line[arrow](130,0)(95,35)
        \Line[arrow](95,35)(130,70)
        \Text(-5,0){$q$}
        \Text(-5,70){$\bar{q}$}
        \Text(65,50){$Z/ \gamma^{*}$}
        \Text(140,0){$\mu^{+}$}
        \Text(140,70){$\mu^{-}$}
\end{axopicture}
    \caption{Drell-Yan \label{fig:bkg1}}
\end{figure}

\begin{figure}
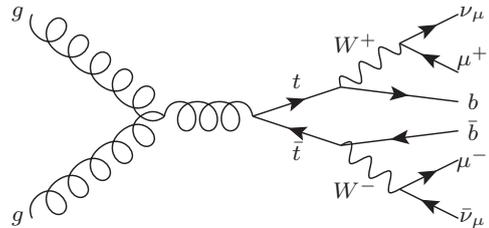

\centering
\begin{axopicture}(130,70)
        \SetScale{1.1}
        \Gluon(-15,0)(30,35){5}{5}
        \Gluon(30,35)(-15,70){5}{5}
        \Gluon(30,35)(60,35){5}{3}
        \Line[arrow](130,30)(90,25)
        \Line[arrow](90,25)(60,35)
        \Line[arrow](60,35)(90,45)
        \Line[arrow](90,45)(130,40)
        \Photon(90,45)(110,60){3}{3}
        \Photon(90,25)(110,10){3}{3}
        \Line[arrow](130,50)(110,60)
        \Line[arrow](110,60)(130,70)
        \Line[arrow](130,0)(110,10)
        \Line[arrow](110,10)(130,20)
        \Text(-20,0){$g$}
        \Text(-20,70){$g$}
        \Text(75,47){$t$}
        \Text(75,23){$\bar{t}$}
        \Text(135,55){$\mu^{+}$}
        \Text(135,70){$\nu_{\mu}$}
        \Text(135,40){$b$}
        \Text(135,20){$\mu^{-}$}
        \Text(135,0){$\bar{\nu}_{\mu}$}
        \Text(135,30){$\bar{b}$}
        \Text(95,10){$W^{-}$}
        \Text(95,60){$W^{+}$}
\end{axopicture}
    \caption{Top pair production \label{fig:bkg2}}
\end{figure}

\begin{figure}
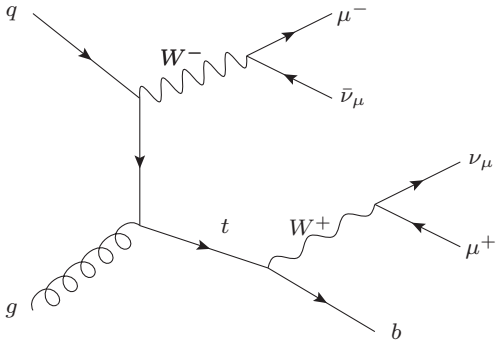

\centering
\begin{axopicture}(210,150)
        \SetScale{0.8}
        \Line[arrow](120,30)(170,00)
        \Line[arrow](60,50)(120,30)
        \Line[arrow](60,110)(60,50)
        \Line[arrow](10,150)(60,110)
        \Gluon(10,10)(60,50){5}{5}
        \Photon(60,110)(110,130){5}{5}
        \Line[arrow](150,110)(110,130)
        \Line[arrow](110,130)(150,150)
        \Photon(120,30)(170,60){3}{3}
        \Line[arrow](210,40)(170,60)
        \Line[arrow](170,60)(210,80)
        \Text(0,10){$g$}
        \Text(0,150){$q$}
        \Text(80,130){$W^{-}$}
        \Text(100,50){$t$}
        \Text(180,0){$b$}
        \Text(220,40){$\mu^{+}$}
        \Text(220,80){$\nu_{\mu}$}
        \Text(160,150){$\mu^{-}$}
        \Text(160,110){$\bar{\nu}_{\mu}$}
        \Text(80,130){$W^{-}$}
        \Text(140,50){$W^{+}$}
\end{axopicture}
    \caption{Single top production \label{fig:bkg3}}
\end{figure}

\begin{figure}
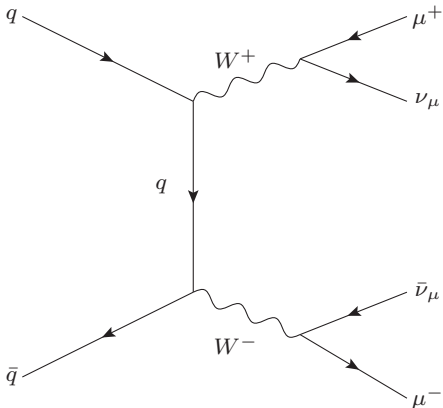

\centering
\begin{axopicture}(200,180)
        \SetScale{0.8}
        \Line[arrow](10,180)(90,140)
        \Line[arrow](90,140)(90,50)
        \Line[arrow](90,50)(10,10)
        \Photon(90,140)(140,160){3}{3}
        \Photon(90,50)(140,30){3}{3}
        \Line[arrow](190,180)(140,160)
        \Line[arrow](140,160)(190,140)
        \Line[arrow](190,50)(140,30)
        \Line[arrow](140,30)(190,0)
        \Text(5,10){$\bar{q}$}
        \Text(5,180){$q$}
        \Text(75,100){$q$}
        \Text(110,160){$W^{+}$}
        \Text(110,25){$W^{-}$}
        \Text(200,180){$\mu^{+}$}
        \Text(200,140){$\nu_{\mu}$}
        \Text(200,50){$\bar{\nu}_{\mu}$}
        \Text(200,0){$\mu^{-}$}
\end{axopicture}
    \caption{Di-Boson production \label{fig:bkg4}}
\end{figure}

We generate the parton level SM background events at leading order in \texttt{Madgraph5}~\cite{Alwall:2011uj}.  These events are then passed to \texttt{Pythia8}~\cite{Sjostrand:2014zea} for the simulation of initial state radiation, parton showering and hadronisation.  Finally, the hadron-level events are passed to \texttt{Delphes3}~\cite{deFavereau:2013fsa} for detector simulation.  We use the 5-flavour \texttt{NNPDF2.3LO} \normalsize ~\cite{Ball:2012cx} parton distribution function via \texttt{LHAPDF6} \cite{Buckley:2014ana} for all background simulations except for di-boson production, for which the 4-flavour \texttt{NNPDF2.3LO} parton distribution function is used.  This choice is made to remove interference in di-boson production, as outlined in more detail later in this section.
There are four significant contributions to the SM background in the $\mu \mu j j$ channel.  These are Drell-Yan ($Z / \gamma^{*} \rightarrow \mu^{+} \mu^{-}$), top pair production ($t \bar{t}$), single top production in association with a $W$ boson ($W t$) and di-boson production ($W^{+} W^{-}$), where top quarks decay leptonically to muons.  An example of the production of each component of the background is shown in Figs. \ref{fig:bkg1} - \ref{fig:bkg4}.  Other sources of background include misidentified muons from $W$+jets, single top production in the $s$ and $t$ channel or multi-jet events.  These form a negligible component of the background in comparison and therefore we treat Drell-Yan, top pair production, single top and di-boson production as the only sources of background.\\

To contribute to the $\mu \mu j j$ signature, Drell-Yan and di-boson production require the addition of at least two jets from initial and final state QCD radiation.  Similarly at least one extra jet must be added to single top production.  To account for this we generate events from processes of a range of different jet multiplicities according to the following definitions:
\begin{align} \label{eq:procdef}
\begin{alignedat}{2}
&\text{DY + 0,1,2,3 jets, \hspace{10pt}}   &&  t \bar{t} \textrm{+ 0,1 jets,}\\
&W t \text{+0,1 jets,} &&W^{+} W^{-} \textrm{+ 0,1,2 jets.}
\end{alignedat}
\end{align}
We include processes with less than two final state jets at parton level to account for the possibility that sufficiently hard jets may be produced by the parton shower algorithm.  We use MLM matching~\cite{Mangano:2006rw} to match the final state partons generated from matrix elements in \texttt{Madgraph5} to those produced by parton showering in \texttt{Pythia8}.  This removes overcounting between multi-jet final states and accounts for the fact that while soft and collinear jets are well described by the parton shower, the matrix elements are more suited to simulating hard and well-separated partons.  MLM matching is implemented in \texttt{Madgraph5} by specifying a nonzero value of the jet cut-off $xqcut$ to be approximately $1/3 $ times a hard scale in the process for each component of the SM background and for each collider.  We confirm our choice of $xqcut$ value in each case by checking that the differential jet rate distributions are smooth and that observables such as the total cross section are insensitive to changes in $xqcut$ about the chosen value.  \\

Interference arises between $WW$ + 2 jets, $Wt$ + 1 jet and $t \bar{t}$ production: all three processes may produce a $WW b \bar{b}$ final state via $t \bar{t}$ production.  Ideally we would simulate all contributons to the $WW b \bar{b}$ final state simultaneously, but this was found to be very computationally difficult and impractical.  Instead, we generate events from each process separately and then combine them to produce the SM background.  This means we must define each process in our simulations in such a way that any overcounting is removed.

A number of methods have been suggested to define $Wt$+1 jet production such
that large contributions from $t \bar{t}$ diagrams are
removed~\cite{Frixione:2008yi}.  We use the diagram removal method as it is
the most straightforward to implement in
\texttt{Madgraph5}~\cite{Demartin:2016axk}.  Let us denote the amplitude for
$Wt$+1 jet by $\mathcal{A}_{Wt}$.  We can write this as $\mathcal{A}_{Wt} =
\mathcal{A}_{1} + \mathcal{A}_{2}$ where $\mathcal{A}_{2}$ is the amplitude
for all diagrams containing $t \bar{t}$ production.  Double counting arises
from the appearance of $|\mathcal{A}_{2}|^{2}$ in $|A_{Wt}|^{2} =
|\mathcal{A}_{1}|^{2} + |\mathcal{A}_{2}|^{2} + 2 {\rm Re}(A_{1}^{\dagger} A_{2})$.
Diagram removal is implemented by setting $\mathcal{A}_{2} = 0$ in our
definition of the $Wt$+1 jet process, which removes the double counting.
Although this method also neglects the interference term $2
{\rm Re}(\mathcal{A}_{1}^{\dagger}\mathcal{A}_{2})$, it has been shown that the
effect of this on observables is moderate and that this method approximates
$Wt$ production well.  We will validate this choice by comparing our
simulations to data in \S\ref{sec:valid}.  The violation of gauge invariance
in the diagram removal method is found to have no observable
effect~\cite{Frixione:2008yi}.

The production of a $W^{+} W^{-} j j$ final state in the di-boson channel
features overcounting as a result of interference with both $t \bar{t}$ and
$Wt$+1 jet production.  This happens only when the two jets originate from $b$
quarks.  In our simulations we remove this interference by treating the $b$
quarks as massive and neglecting them from the definitions of the proton and
jets i.e.\ by working in a 4-flavour scheme.  This is the method used by ATLAS
in their simulations at $13$ TeV~\cite{ATL-PHYS-PUB-2017-005}.  4-flavour
parton distribution functions are used.  This removes all $t \bar{t}$ and $W
t$+1 jet production from the di-boson channel, but also neglects processes
with initial and final state $b$ quarks which do contribute to di-boson
production.  A study of how well this 4-flavour scheme approximates the full
di-boson production cross section at centre of mass energies $\sqrt{s} =
14,100$ TeV was undertaken in Ref.~\cite{Mangano:2017tke} by comparing the leading order cross sections of di-boson production in the 4 and 5-flavour schemes, where it was found that the difference is negligible at $14$ TeV and $\sim 5 \%$ at $100$ TeV.\\

To produce an accurate simulation of the SM background at each future collider, we use \texttt{Delphes3} to simulate the response of the detectors and the decay of short-lived particles.  Jets are clustered using the anti-$k_{T}$ clustering algorithm~\cite{Cacciari:2008gp} with jet radius $R=0.4$.  This value is chosen from the ATLAS analysis at 13 TeV~\cite{Aaboud:2016qeg} to reproduce their analysis as closely as possible.  To mimic the response of different detectors at each future collider we specify detector configurations as follows.  The ATLAS configuration is used in all simulations at 13 TeV.  At 14 TeV and 27 TeV we use the \texttt{Delphes3} HL-LHC configuration designed to reproduce the average response of the ATLAS and CMS detectors at high energies and luminosities.  Similarly in our simulations of the 100 TeV FCC-hh we use the FCC-hh configuration provided by \texttt{Delphes3}.  We maintain the default settings in our simulations except in the case of muon isolation.  Muon isolation is defined by finding the sum of the transverse momentum $p_{T}$ of all objects within a cone of radius $R^\text{max}$ around a muon, excluding the $p_{T}$ of the muon itself.  If the sum satisfies $p_{T}^{sum} < p_{T}^\text{max}$ for fixed $p_{T}^\text{max}$, the muon is considered isolated.  At 13 TeV and 14 TeV we select only isolated muons with $p_{T}^\text{max} = 0.2$ GeV and $R^\text{max} = 0.2$, choosing these parameters to reproduce the 13 TeV ATLAS analysis.  At 27 TeV and 100 TeV we make no selection on the muon isolation criteria, following the same reasoning as in~\cite{Jamin:2019mqx}.  This choice is made because the overall normalisation of the SM background is found to be very dependent on the muon isolation criteria and the specific selection made will likely vary in different future analyses.  Relative to our simulations, any selection on muon isolation at future experiments will only reduce the SM background producing a better sensitivity to the LQ signal.\\

We are interested in the search for TeV-scale LQs, which are expected to
manifest as a resonance at high $m_\text{min}(\mu,j)$.   Producing a large
number of events in the tail of the $m_\text{min}(\mu,j)$ distribution is
therefore necessary to achieve good statistics in this region.  We find that
binning the generation of events in $m_\text{min}(\mu, j)$ at parton-level or
in parameters such as the dimuon invariant mass $M_{\mu \mu}$ and $H_{T} =
p_{T}^{j_{1}} + p_{T}^{j_{2}}$ is inefficient for producing a sufficient
number of  tail
events.  Instead we reweight the generation of each event $x$ by applying a
bias $b(x) \propto s(x)^{5}$.  For each SM background process $s(x)$ is
defined at parton-level as the invariant mass of the final state muons and
jets, where we only include the minimum number of jets in the multi-jet
process definitions of Eq.~\ref{eq:procdef}, accounting for jets originating
from top quarks.  For example, for Drell-Yan we define $s(x)$ as the invariant
mass of the di-muon final state.  All physical observables and distributions
shown in this paper have been obtained by unweighting the events after parton
showering and detector simulation, in order to remove the effect of this bias.

\subsection{Validation}
\label{sec:valid}
We first validate our methods by simulating the SM background at
$\sqrt{s} = 13$ TeV for an integrated luminosity $\mathcal{L} = 3.2$ fb$^{-1}$ and comparing with the
ATLAS search for second generation LQs at the same centre of mass energy and
integrated luminosity~\cite{Aaboud:2016qeg}.  We compare our simulations to
the ATLAS data in two regions of phase space: the preselection region and the
signal region.  Both are defined by cuts on $p_{T}$, $|\eta|$ and $\Delta R$
designed to increase the significance of a LQ signal above the SM background
and are  summarised in Table~\ref{table:cuts1}.  All jet cuts are placed on
the two hardest jets in the event denoted by $j_{1}$, $j_{2}$.  The signal
region is subject to further cuts on $S_{T} = p_{T}^{\mu_{1}} +
p_{T}^{\mu_{2}}  + p_{T}^{j_{1}}  + p_{T}^{j_{2}}$ and $M_{\mu \mu}$.  These
significantly reduce features of the SM background due to soft jets and $W$
and $Z$ boson resonances.  In both the signal and preselection regions we also reject muons falling in the range $1.01 < |\eta| < 1.1$ as specified by the ATLAS analysis to avoid potential $p_{T}$ mismeasurement in this range.  A preliminary selection on muon isolation is made at the level of detector simulation as outlined in \S\ref{sec:methods}.

\begin{table}[h]
\begin{center}
\begin{tabular}{ |c|c|c|c|c| }
 \hline
 Region& $p_{T}^{j}$ &$p_{T}^{\mu}$  & $|\eta_{\mu}|$ & $|\eta_{j}|$\\
 \hline 
 Preselection &  $>50$ & $>40$ &  $<2.5$ & $< 2.8$ \\
 Signal   & $>50$ &$ >40$ &  $<2.5$ & $< 2.8$ \\
 \hline \hline
 Region& $\Delta R_{\mu j}$ & $\Delta R_{\mu \mu}$ & $M_{\mu \mu}$  & $S_{T}$ \\
 \hline
 Preselection & $>0.4$ &$>0.3$ & & \\
 Signal    & $>0.4$ &$>0.3$  &$>130$ & $> 600$  \\
 \hline
\end{tabular}
	\caption{Phase space cuts defining the preselection and signal regions at $\sqrt{s} = 13$ TeV.  All cuts are applied in the analysis after parton showering and detector simulation.  Cuts on $p_{T}$, $M_{\mu \mu}$ and $S_{T}$ are in units of GeV.} \label{table:cuts1}
\end{center}
\end{table}
To efficiently simulate events in these regions of\\ phase space, we generate events subject to a subset of the phase space cuts.  These are applied at parton-level in the \texttt{Madgraph5} run card and summarised in Table~\ref{table:cuts2}.  We will refer to this subset as generator cuts.  The jet cut off $xqcut$ required for MLM matching is found for each process in the presence of the generator cuts, as outlined in \S\ref{sec:methods}.  Note however that we set the parameter \texttt{auto$\_$ptj$\_$mjj = True} for DY and di-boson production, allowing the jet matching procedure to automatically set the cuts on $p_{T}^{j}$ and $M_{j_{1} j_{2}}$ equal to the chosen value of $xqcut$.  We use $xqcut = 30,60,60,30$ GeV for DY, top pair, single top and di-boson production, respectively.

\begin{table}[h]
\begin{center}
\begin{tabular}{ |c|c|c|c| }
 \hline
	$p_{T}^{j_{1}}$ &$p_{T}^{\mu}$ &  $M_{\mu \mu}$ (pres.) &  $M_{\mu \mu}$ (sig.) \\
 \hline 
 $>35$ & $>30$ &  $> 20$  &  $> 120$   \\
 \hline \hline
  $|\eta_{\mu}|$ & $|\eta_{j}|$& $\Delta R_{\mu j}$ & $\Delta R_{\mu \mu}$ \\
 \hline 
$<2.5$ & $< 2.8$ & $>0.4$ &$>0.3$ \\

 \hline
\end{tabular}
\caption{Cuts applied at parton-level to efficiently simulate events at $\sqrt{s} = 13$ TeV in the preselection region and signal region.} \label{table:cuts2}
\end{center}
\end{table}
 \begin{figure}[h]
  \begin{center}
    \unitlength=\textwidth
    {\includegraphics[width=0.45 \textwidth]{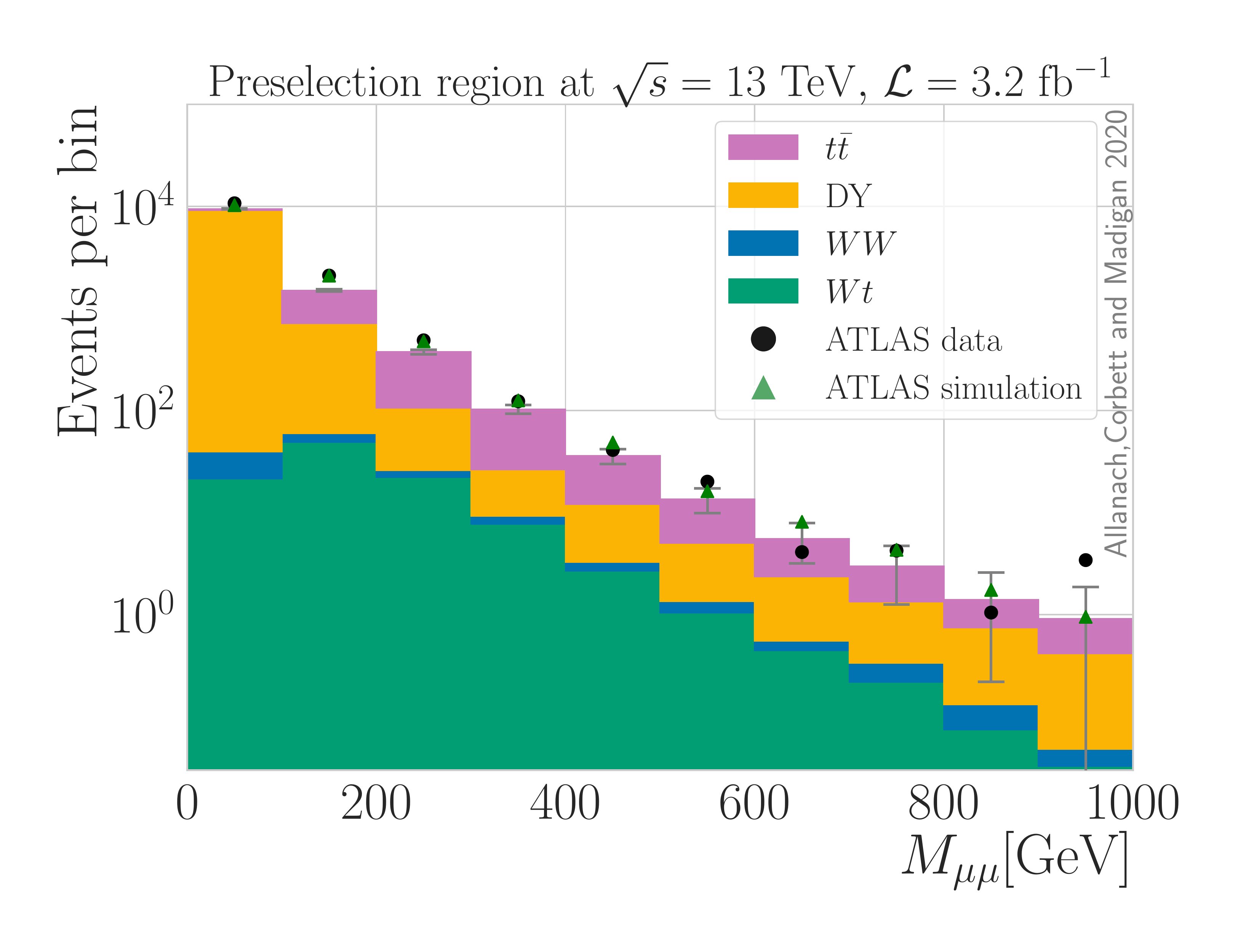}}
  \end{center}
  \caption{\label{fig:13TeVpres}   Validation plot showing our simulations of the $M_{\mu \mu}$ distribution of the SM backgrounds in the search for the pair production of second generation LQs in the $\mu \mu j j$ channel at $\sqrt{s} = 13$ TeV, $\mathcal{L} = 3.2$ fb$^{-1}$ in the preselection region.  We compare our simulations to the ATLAS simulations and data for validation.}
  \end{figure}

  \begin{figure}[h]
  \begin{center}
    \unitlength=\textwidth
    {\includegraphics[width=0.45 \textwidth]{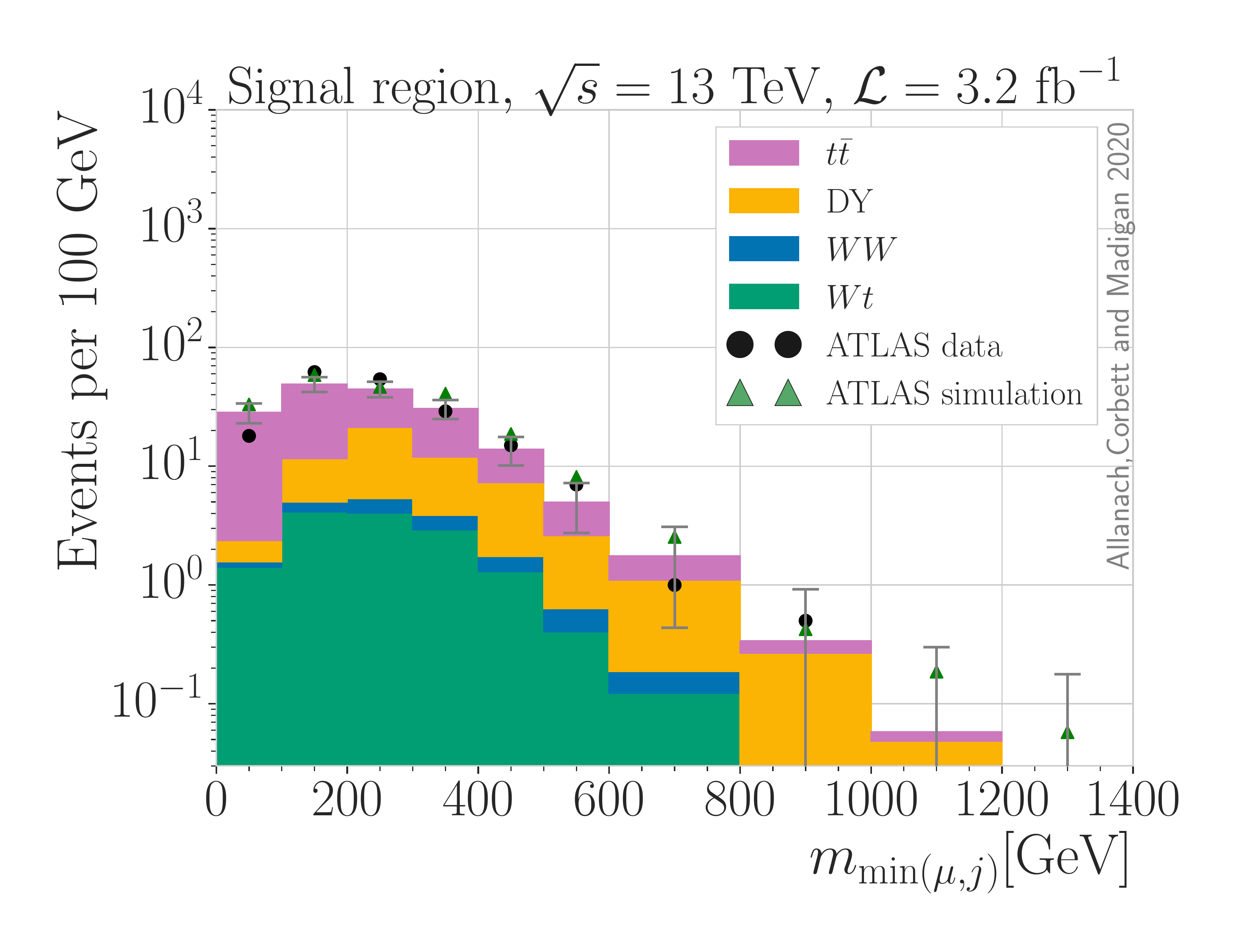}}
  \end{center}
          \caption{\label{fig:13TeVback}   Validation plot showing our simulations of the $m_\text{{min}}(\mu, j)$ distribution of the SM backgrounds in the search for the pair production of second generation LQs in the $\mu \mu j j$ channel at $\sqrt{s} = 13$ TeV, $\mathcal{L} = 3.2$ fb$^{-1}$ in the signal region.  We compare our simulations to the ATLAS simulations and data for validation.}
  \end{figure}

Fig. \ref{fig:13TeVpres} shows the distribution of preselection events in the parameter $M_{\mu \mu}$.    The Monte Carlo error on each bin is shown in grey and is computed from the Monte Carlo event weights $w_{i}$ by $\mathrm{Err}_{i} = \sqrt{\sum w_{i}^{2}}$.  Systematic uncertainties are not included.  Our simulations are not in perfect agreement with the ATLAS data (simulations) shown in black (green).  This is expected because we generate all events at leading order, and the dominant process in this region is $t \bar{t}$ production which has large NLO corrections.  However, our simulations provide a good estimation of the order of magnitude of the SM background in each bin.

Our methods are further validated in Fig. \ref{fig:13TeVback} which shows the distribution of signal region events in the parameter $m_\text{min}(\mu,j)$.  As in the preselection region, we underestimate the SM background slightly by working only at leading order.  However, compared to the preselection region, this provides a less fair comparison as normalisation factors have been applied to rescale to the ATLAS simulations in the signal region.  Overall we take this comparison as a validation of our methods for simulating the SM background.

\subsection{Future collider backgrounds}
\label{sec:FCback}
\begin{table*}
\begin{center}
\begin{tabular}{ |c|c|c|c|c|c|c| }
 \hline
Collider& $p_{T}^{j} $ (GeV)  &$p_{T}^{\mu} $ (GeV) & $|\eta_{\mu}| $ & $|\eta_{j}| $  & $M_{\mu \mu} $ (GeV) & $S_{T} $ (GeV)\\
 \hline \hline
	\small{LHC} &  $>50$ & $>40$ &  $<2.5$ & $<2.8$ & $>130$ & $>600$  \\
	\small{HL-LHC}   & $>50$ &$ >40$ &  $<2.5$ & $ <2.8$  &$>130$ & $ >600$  \\
	\small{HE-LHC} &  $>100$ & $>80$ &  $<4.0$ & $ <4.0$  & $>260$ & $>1200$  \\
	\small{FCC-hh}   & $>400$ &$ >320$ &  $<4.0$ & $< 4.0$  &$>1000$ & $ >4000$  \\
 \hline
\end{tabular}
\caption{Phase space cuts defining the signal regions in simulations of the 13 TeV LHC and future colliders.  All cuts are applied in the analysis after parton showering and detector simulation.} \label{table:cuts3}
\end{center}
\end{table*}

We generate the SM background at the LHC with the full Run II integrated
luminosity (13 TeV, $140$ fb$^{-1}$) and at the three future colliders
in Table~\ref{table:colliders}.  We define the signal region at each collider by a
set of phase space cuts based on the 13 TeV ATLAS analysis as follows.  All
angular separations $\Delta R$ are kept unchanged.  For the LHC and HL-LHC we
keep the same cuts on $|\eta|$, while at the HE-LHC and FCC-hh these are
increased to $|\eta| < 4$ to allow for a difference in detector topologies at
the future colliders.  Cuts with dimensions of energy ($p_{T}$, $S_{T}$,
$M_{\mu \mu}$) are kept constant for the HL-LHC and scaled up by approximately
the ratio of centre of mass energies $\sqrt{s} /( 13$  TeV) for the HE-LHC and
FCC-hh simulations.  These signal region cuts are summarised in
Table~\ref{table:cuts3}.
As in our 13 TeV simulations, we generate events subject to generator cuts applied at parton-level in the \texttt{Madgraph5} run card, summarised in Table~\ref{table:cuts4}.  Table~\ref{table:xqcut} specifies the values of $xqcut$ used in MLM matching for each component of the SM background.
\begin{table}[h]
\begin{center}
\begin{tabular}{|c|c|c|c|c|c|c|}
 \hline
Collider&  $p_{T}^{j_{1}}$ &$p_{T}^{\mu}$  & $|\eta_{\mu}|$ & $|\eta_{j}|$ &  $M_{\mu \mu}$\\
 \hline \hline
LHC &  $>35$ & $>30$ &  $<2.5$ & $< 2.8$ & $>130$  \\
HL-LHC   & $>35$ &$ >30$ &  $<2.5$ & $< 2.8$ &$>130$  \\
HE-LHC &  $>85$ & $>60$ &  $<4.0$ & $< 4.0$ & $>200$ \\
FCC-hh   & $>380$ &$ >300$ &  $<4.0$ & $< 4.0$ &$>900$  \\
 \hline
\end{tabular}
	\caption{Phase space cuts applied at parton-level in \texttt{Madgraph5} to efficiently simulate events in the signal region for the 13 TeV LHC and future colliders.  Cuts on $p_{T}$ and $M_{\mu \mu}$ are in units of GeV.} \label{table:cuts4}
\end{center}
\end{table}
\begin{table}[h]
\begin{center}
\begin{tabular}{|c|c|c|c|c|}
 \hline
Collider& DY  & $t \bar{t}$  & $Wt$ & $W^{+} W^{-}$\\
 \hline \hline
LHC &  $30$ & $60$ &  $30$ &  $30$  \\
HL-LHC   & $30$ &$60$ &  $30$ &  $30$  \\
HE-LHC &  $45$ & $120$ &  $120$  &  $60$ \\
FCC-hh   & $90$ &$300$ &  $120$  &  $120$ \\
 \hline
\end{tabular}
\caption{Values of the jet cut-off parameter $xqcut$ in GeV used in MLM matching of multi-jet events at the 13 TeV LHC and future colliders.  All jet matching parameters are found in the presence of the generator cuts summarised in Table~\ref{table:cuts4}. \label{table:xqcut}}
\end{center}
\end{table}

\begin{figure*}[htp]
   \subfloat[]{
      \includegraphics[width=.47\textwidth]{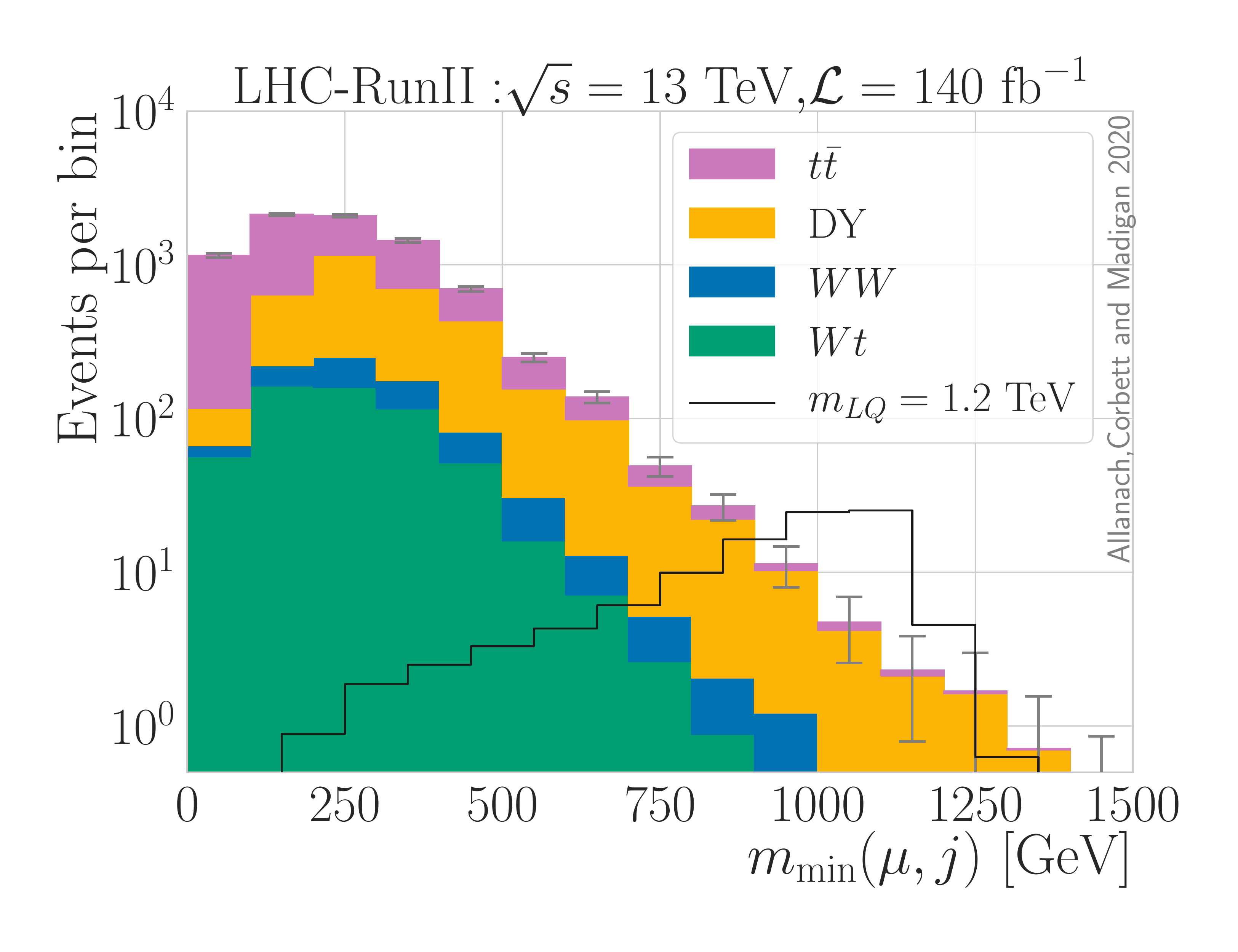}}
   \subfloat[]{
      \includegraphics[width=.47\textwidth]{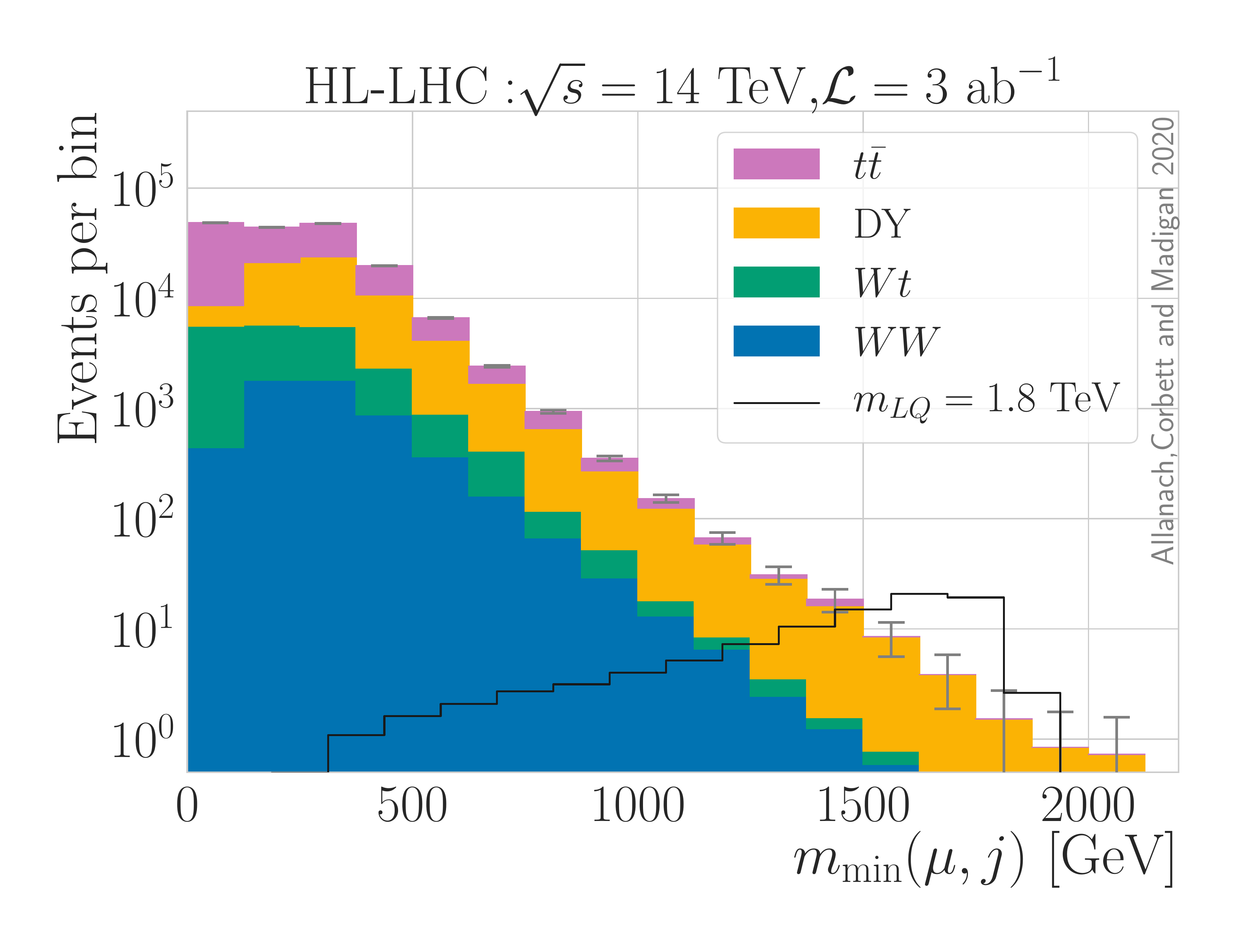}}
        \caption{Predicted $m_\text{min}(\mu,j)$ distribution of the SM background and an example of a LQ signal at the 13 TeV LHC with full Run II integrated luminosity of $\mathcal{L} = 140$ fb$^{-1}$ (a) and at the HL-LHC (b).  The LQ signals correspond to $m_\text{{LQ}} = 1.2$ TeV and $m_\text{{LQ}} = 1.8$ TeV respectively, with couplings chosen to fit the NCBAs as outlined in \S\ref{sec:signal}, and have statistical significances of $5 \sigma$ and $7 \sigma$ relative to the SM background. \label{fig:13TeVHLback}}
\end{figure*}

\begin{figure*}[htp]
   \subfloat[]{
      \includegraphics[width=.47\textwidth]{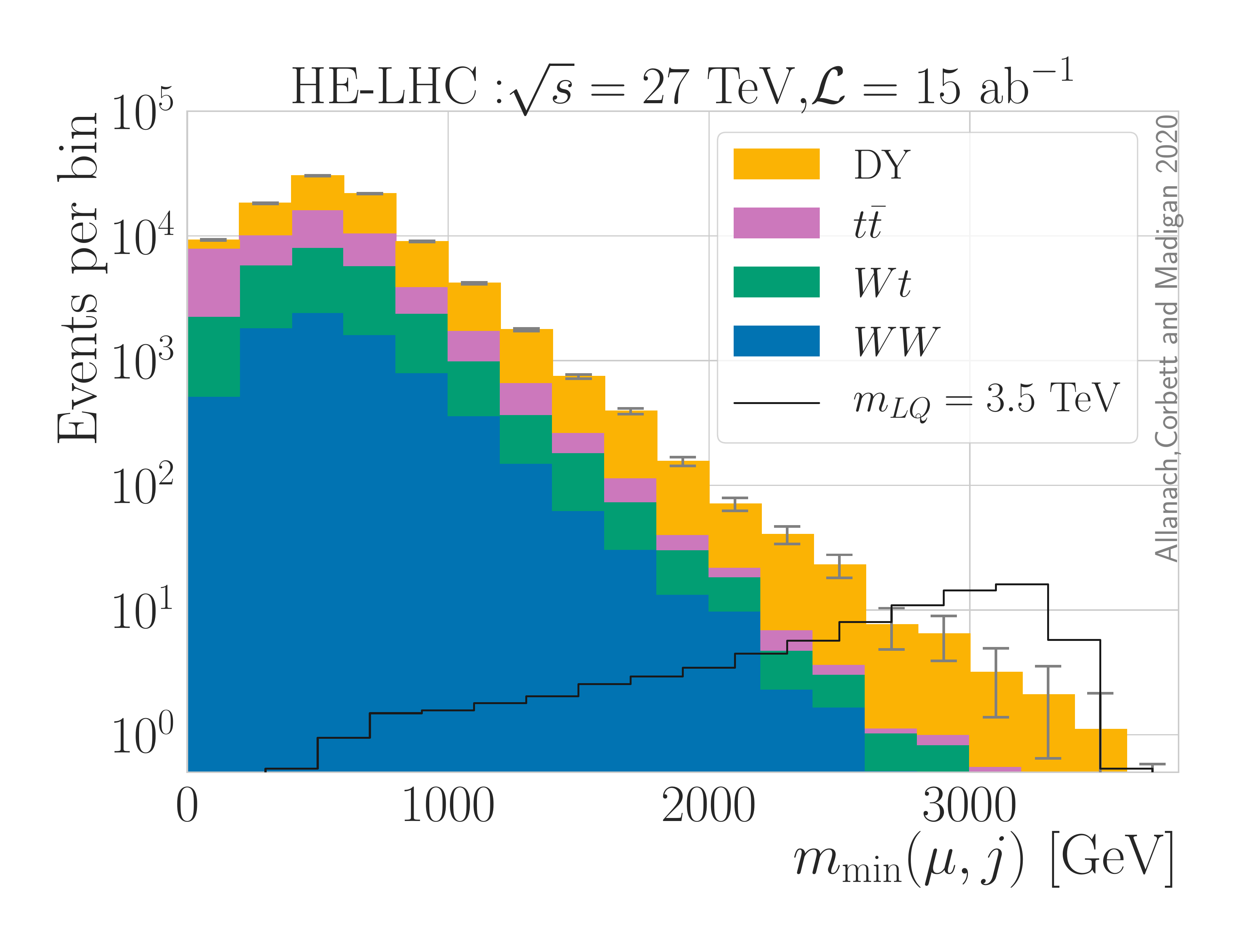}}
   \subfloat[]{
      \includegraphics[width=.47\textwidth]{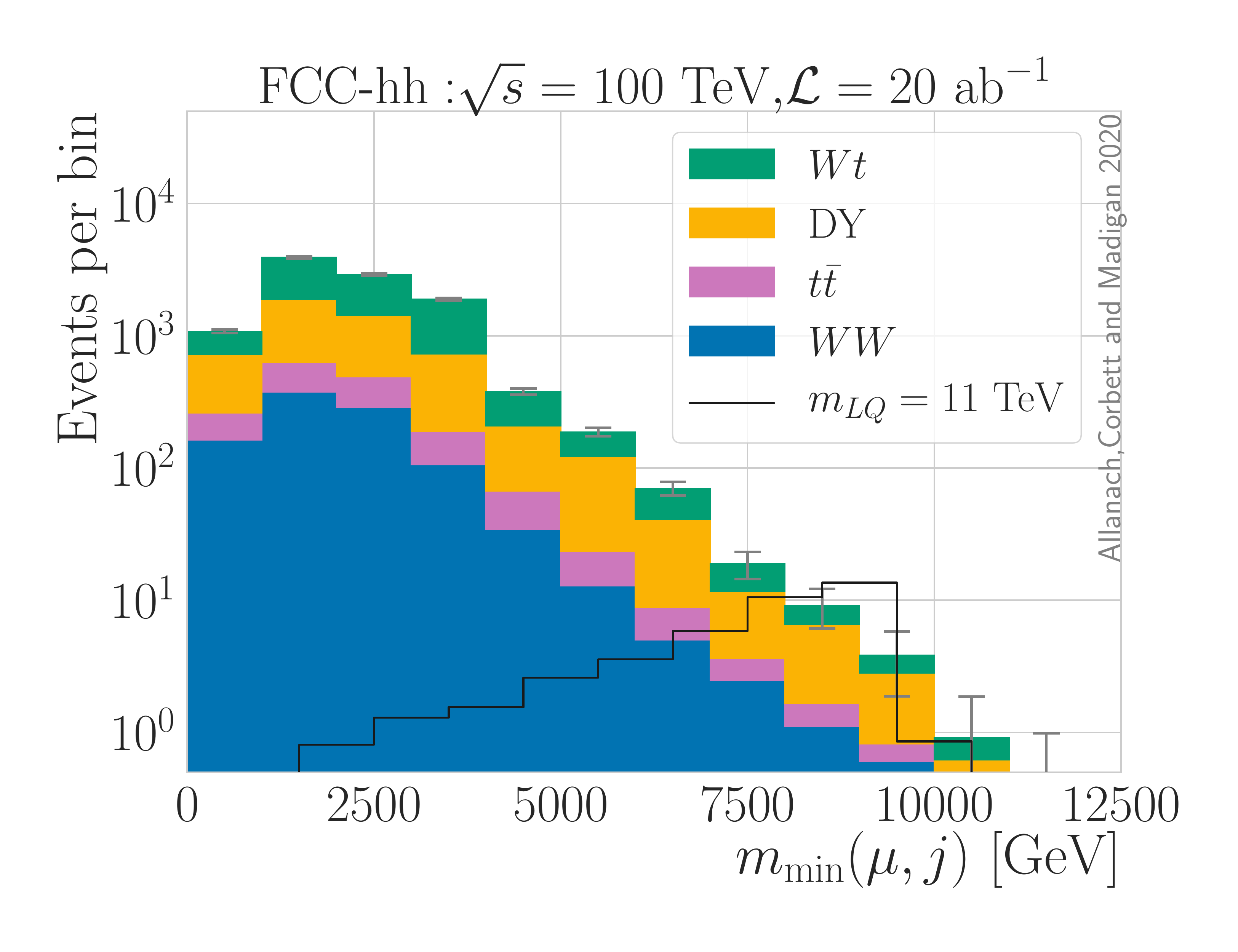}}
        \caption{Predicted $m_\text{min}(\mu,j)$ distribution of the SM background and an example of a LQ signal at the HE-LHC (a) and at the FCC-hh (b).  The LQ signals correspond to $m_\text{{LQ}} = 3.5$ TeV and $m_\text{{LQ}} = 11$ TeV respectively, with couplings chosen to fit the NCBAs as outlined in \S\ref{sec:signal}, and have statistical significances of $6 \sigma$ and $3 \sigma$ relative to the SM background. \label{fig:27TeVback}}
\end{figure*}

Figs. \ref{fig:13TeVHLback} and \ref{fig:27TeVback} show the resulting distributions of the SM background at the LHC, HL-LHC, HE-LHC and FCC-hh respectively.  As before the Monte Carlo error is shown in grey and systematic uncertainties are not included.

\section{Sensitivity \label{sec:sens}}
\subsection{Signal simulations} \label{sec:signal}
To find the significance of a LQ at mass $m_\text{LQ}$ relative to these
backgrounds, we simulate the distribution of a LQ resonance in
$m_\text{min}(\mu, j)$.  We simulate LQ pair production and decay into a $\mu
\mu j j$ final state at leading order\footnote{The leptoquark
pair production signal at
  hadron colliders was calculated some time ago at leading
  order~\cite{PhysRev.124.1577,Blumlein:1996qp}.
  Next-to-leading order
  effects~\cite{Kramer:1997hh,Kramer:2004df} and parton shower
  effects~\cite{Mandal:2015lca} on the signal have also been studied.} in \texttt{Madgraph5}.  We work in a
5-flavour scheme using the 5-flavour \texttt{NNPDF2.3LO}~\cite{Ball:2012cx}
parton distribution function, and correct for parton showering and detector
effects using \texttt{Pythia8} and \texttt{Delphes3} as before.  We simulate
LQs from the $S_{3}$ LQ model provided by\footnote{A link to the \texttt{UFO} model files can be found within this reference.}~\cite{Dorsner:2018ynv}.

We specify LQ couplings as follows.  For each $m_\text{LQ}$ the product of
couplings $|(Y_{de})_{32} (Y^{*}_{de})_{22}|$ is fixed by fits to the NCBAs as given in Eq.~\ref{eftMatch}.  We choose
$(Y_{de})_{22} = (Y_{de})_{32}$ and set all other $(Y_{de})_{ij}$ to zero.  This couples the LQ to $b_{L} \mu_{L}$ and $s_{L} \mu_{L}$ pairs as required by the NCBAs.  Following the conventions of Ref.~\cite{Dorsner:2016wpm}, all CKM and PMNS mixing occurs within the up and neutrino sector respectively i.e.\ we set $V_{d_{L}} = V_{e_{L}} = I$ in Eq.~\ref{unPrimedS3}.  Our choice of couplings then corresponds to setting $(Y_{L})_{22} = (Y_{L})_{32} \neq 0$, inducing further couplings of the LQ to $\mathbf{u}_{L} \mu_{L}$, $b_{L} \boldsymbol{\nu}_{L}$, $s_{L} \boldsymbol{\nu}_{L}$ and $\mathbf{u}_{L} \boldsymbol{\nu}_{L}$ pairs where $\mathbf{u}_{L}$ and $\boldsymbol{\nu}_{L}$ denote the vectors of left-handed up-type quarks and neutrinos respectively.  These include CKM suppressed couplings to $u_{L} \mu_{L}$ and $c_{L} \mu_{L}$ which will contribute to the muon-jet decay channel of the LQ, increasing the number of events in the $\mu \mu j j$ signal.  Similarly, we account for these additional CKM and PMNS suppressed couplings in calculating the theory predictions for $\sigma \times \text{BR}$, taking the central values of $V_{\text{CKM}}$ and $U_{\text{PMNS}}$ from \cite{Tanabashi:2018oca} assuming normal ordering of neutrino masses.

As outlined in \S\ref{sec:valid} and \S\ref{sec:FCback} we generate
events subject to the generator cuts in Table~\ref{table:cuts4} at
parton-level, applying the full set of signal region phase space cuts from Table~\ref{table:cuts3} in the analyses.  Examples of the predicted distributions of
LQ events at each future collider are included in Figs.
\ref{fig:13TeVHLback} and \ref{fig:27TeVback}.

\subsection{Statistics}
Each experiment consists of measurements of events in $N$ bins of a histogram, denoted by $n_{i}$ where $i=1,\ldots,N$.  We find the expected number of background events $b_{i}$ and signal events $s_{i}$ in bin $i$ from our Monte Carlo simulations, and parametrise the signal present in our data sample by the signal strength $\mu \in [0,1]$.  The likelihood is defined by taking the product of Poisson probabilities in each bin
\begin{align}
L(\mu, \theta) = \prod_{i=1}^{N} \frac{(\mu s_{i} + b_{i})^{n_{i}}}{n_{i}!} e^{-(\mu s_{i} + b_{i})},
\end{align}
where $\theta$ denotes all nuisance parameters.  Defining the profile likelihood ratio as $\lambda(\mu) = L(\mu, \hat{\hat{\theta}})/L(\hat{\mu}, \hat{\theta})$ where $\hat{\mu}$ and $\hat{\theta}$ are the maximum likelihood estimates of $\mu$ and $\theta$, and $\hat{\hat{\theta}}$ is found by maximising the likelihood with fixed $\mu$.

Suppose the measured data $n_{i}$ shows no fluctuations above the SM background $b_{i}$.  To set exclusion limits on $\sigma \times \text{BR}$ we test the $b+\mu s$ hypothesis and find the maximum value of $\mu$ compatible with the data.  We quantify compatibility by computing the $p$-value from the modified frequentist $\text{CL}_{s}$ method~\cite{Junk:1999kv} and the test statistic $q_{\mu}$ defined by
\begin{align} \label{eq:qmu}
	q_{\mu} = \left\{
\begin{array}{l}
	-2 \textrm{ln} \lambda(\mu)  \hspace{10pt} \hat{\mu} \leq \mu\\
	0 \hspace{10pt} \hspace{38pt} \hat{\mu} > \mu.
\end{array}
\right.
\end{align}
The upper limit at $95 \%$ CL on $\mu$ is then given by the value of $\mu$ at which $\text{CL}_{s} = 0.05$.  We compute the $\text{CL}_{s}$ values using \texttt{pyhf}~\cite{lukas_2019_3245478}, a Python implementation of \texttt{HistFactory}~\cite{Cranmer:2012sba}.  By comparison with the theoretical predictions, any $(\sigma \times \text{BR})_{\text{theory}} > (\sigma \times \text{BR})_{\text{lim}}$ can then be excluded.  This determines the mass sensitivity i.e.\ the maximum $m_\text{LQ}$ that could be excluded at $95 \%$ CL at each future collider.

Alternatively, suppose an excess of events is seen in the data $n_{i}$.  To find the significance of such an observation we test the compatibility of the background-only hypothesis $\mu = 0$ with the data.  The test statistic $q_{0}$ is defined by
\begin{align} \label{eq:qmu0}
	q_{0} = \left\{
\begin{array}{l}
	-2 \textrm{ln} \lambda(0)  \hspace{10pt} \hat{\mu} \geq 0\\
	0 \hspace{10pt} \hspace{38pt} \hat{\mu} < 0.
\end{array}
\right.
\end{align}
The discovery reach of each future collider is found by determining, for each $m_\text{LQ}$ of interest, the integrated luminosity $\mathcal{L}$ required for a $p$-value of $\text{CL}_{s} = 2.9 \times 10^{-7}$ or equivalently a  statistical signifiance of $5 \sigma$.

In determining the discovery and exclusion sensitivities, we work in the large sample approximation and use the Asimov data set to calculate the median $\text{CL}_{s}$~\cite{Cowan:2010js}.

 \begin{figure}[H]
  \begin{center}
    \unitlength=\textwidth
    {\includegraphics[width=0.48 \textwidth]{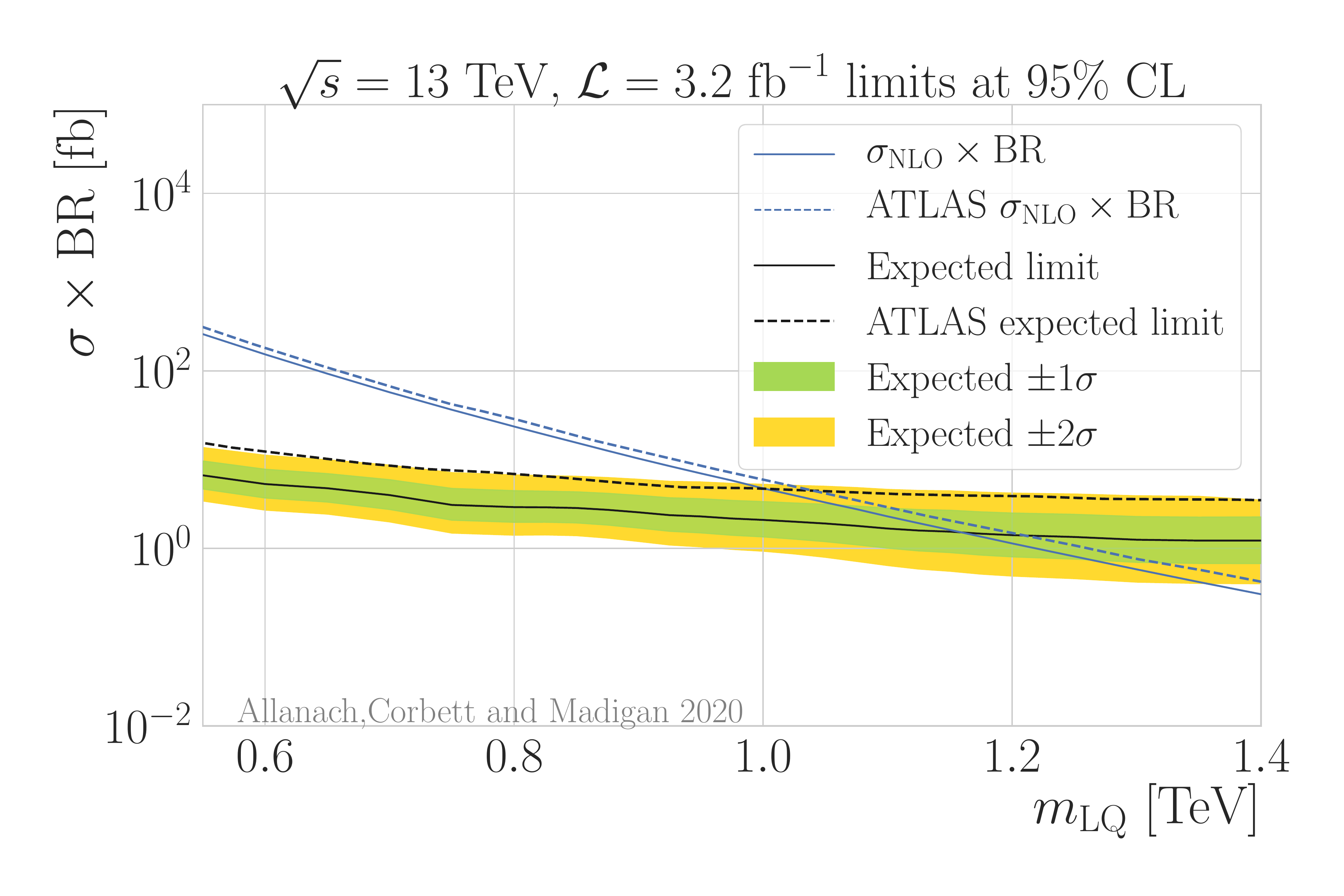}}
  \end{center}
  \caption{\label{fig:13TeVlimits} Validation plot comparing our expected limits at $95 \%$ CL on $\sigma \times \text{BR}$ for LQ pair production and decay into a $\mu \mu j j$ final state at $\sqrt{s} = 13$ TeV, $\mathcal{L} = 3.2$ fb$^{-1}$ to the expected limits obtained by ATLAS.}
  \end{figure}

We validate this method by using the signal region data generated at $\sqrt{s}
= 13$ TeV, $\mathcal{L} = 3.2$ fb$^{-1}$ in \ref{fig:13TeVback} to place
limits on LQs in the range $m_{\rm LQ} \in [500,1400]$ GeV.  The resulting
limits on $\sigma \times \text{BR}$ as a function of $m_{\text LQ}$ are shown in
Fig. \ref{fig:13TeVlimits}, excluding LQ masses up to approximately 1.15 TeV.
This limit is compared to the exclusion limits found by ATLAS, shown by the
black dashed curve, indicating sensitivity to LQ masses up to 1.05 TeV.  Note
that for the purposes of this comparison only we generate events and compute
the $\sigma \times \text{BR}$ from a model of second generation LQs decaying
into a $\mu^{-} \mu^{+} c \bar{c}$ final state with coupling $y_{\mu c} =
\sqrt{0.01 \times 4 \pi \alpha_{em}}$ from the minimal
Buchm{\"u}ller-R{\"u}ckl-Wyler model~\cite{Buchmuller:1986zs}, following the
ATLAS 13 TeV analysis.  All other LQ events and values of $\sigma \times
\text{BR}$ in this paper are found as outlined in \S\ref{sec:signal} and
\S\ref{sec:wide}.

This shows that our methods have slightly overestimated the sensitivity to LQ\@.  This is to be expected from the fact that we have underestimated the SM background and do not include systematic uncertainties in setting limits.  However, as an estimate of the sensitivity this is a good approximation, and so we take this comparison as a validation of our methods and proceed by using the same methods for future colliders.

\subsection{Future colliders}
\begin{figure*}[htp]
   \subfloat[]{
      \includegraphics[width=.48\textwidth]{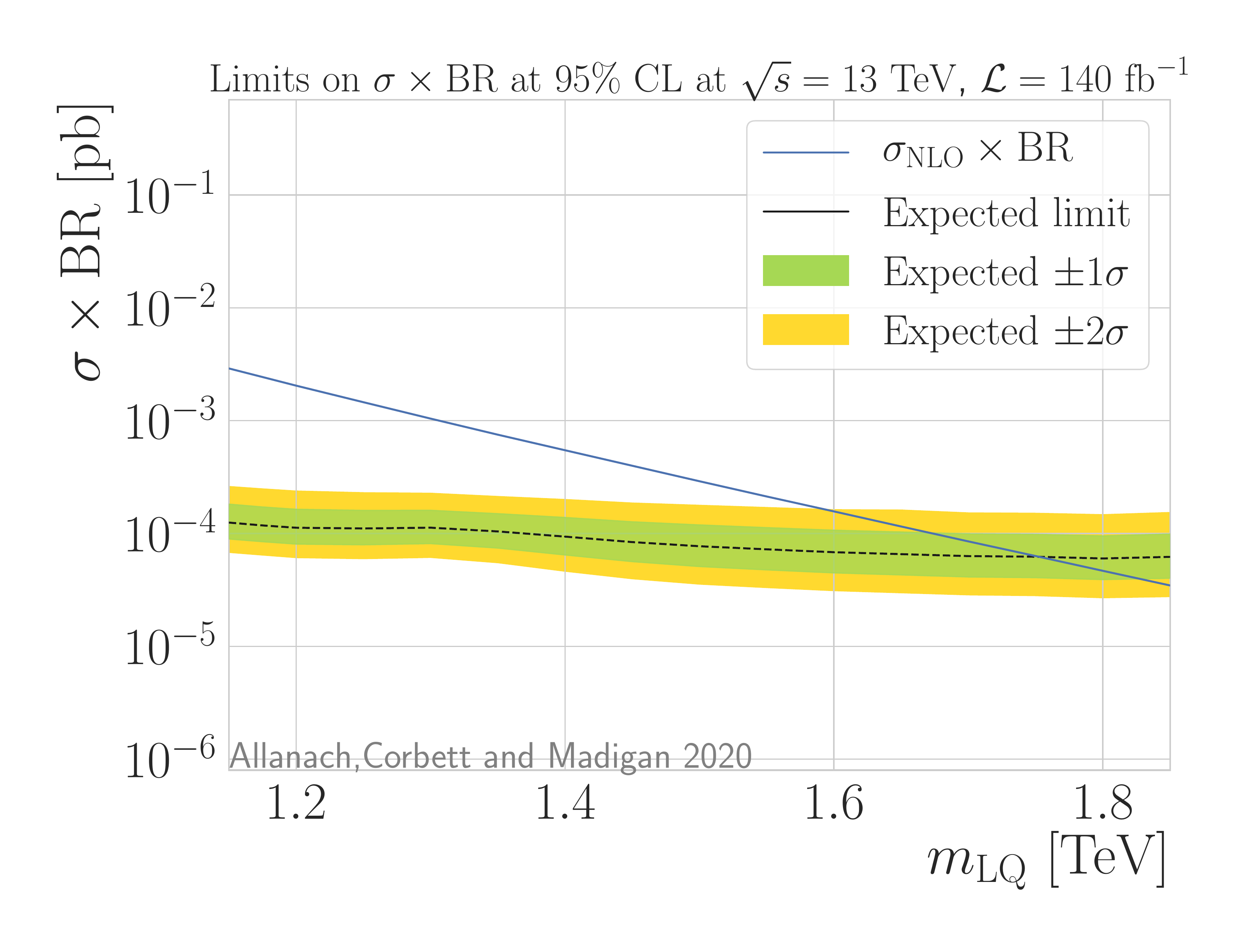}}
~
   \subfloat[]{
      \includegraphics[width=.48\textwidth]{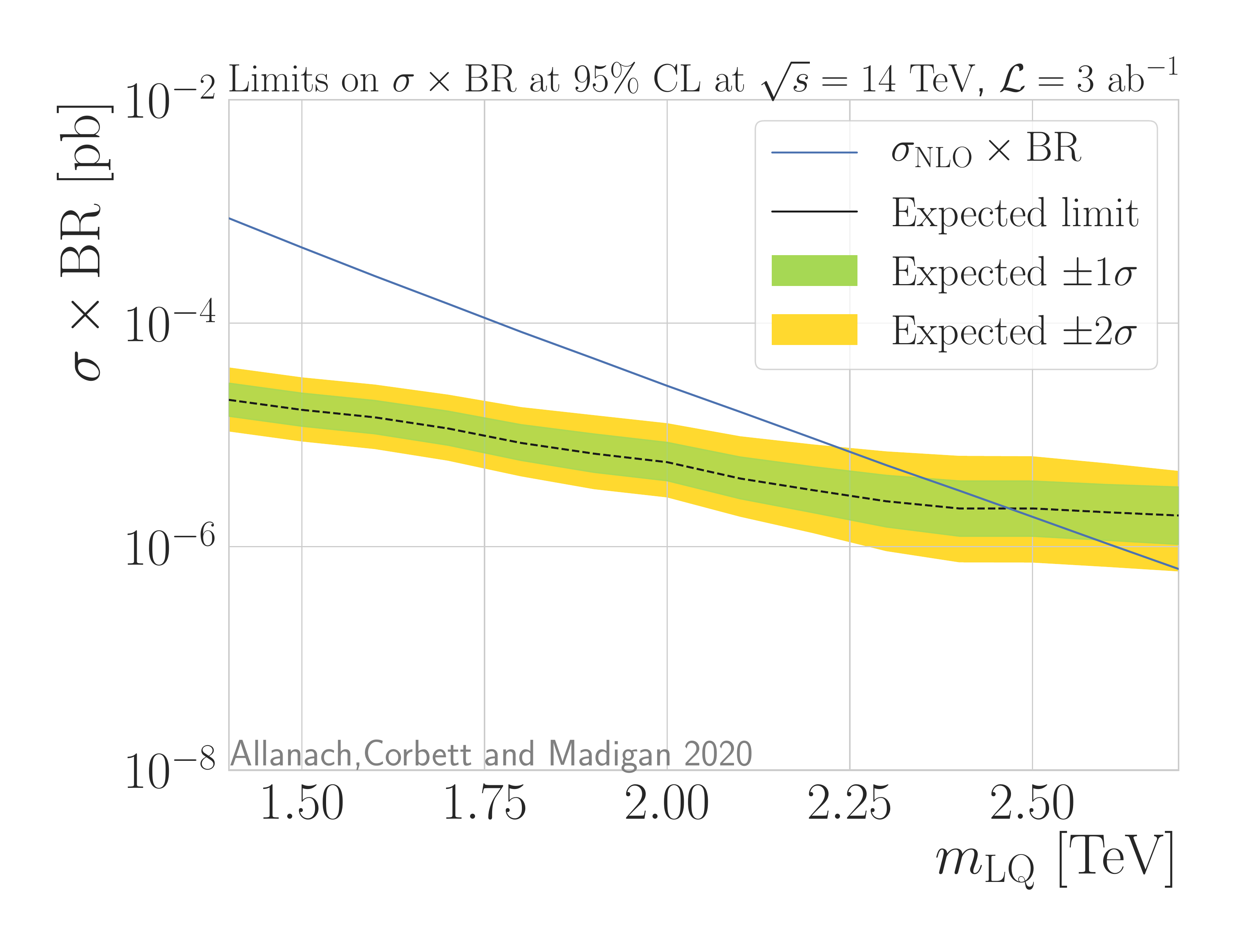}}
        \caption{Expected limits at $95 \%$ CL on $\sigma \times \text{BR}$ for the pair production of LQs and decay into a $\mu \mu j j$ final state at the LHC with full Run II integrated luminosity (a) and the HL-LHC (b).  Theory curves $\sigma_{\text{NLO}} \times \text{BR}$ are calculated for narrow width LQs with couplings chosen to fit the NCBAs.}
        \label{fig:limits1}
\end{figure*}

\begin{figure*}[htp]
   \subfloat[]{
      \includegraphics[width=.48\textwidth]{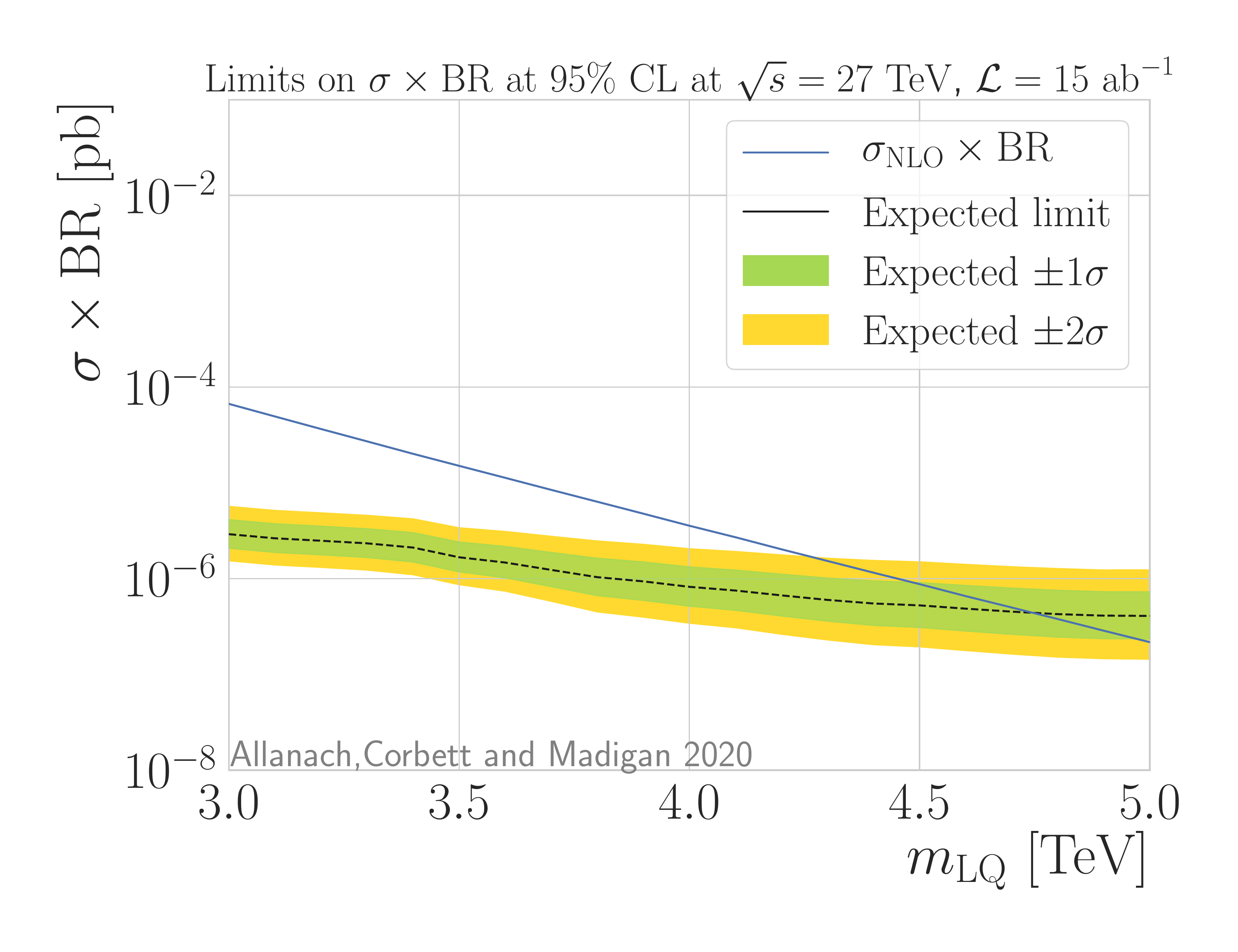}}
~
   \subfloat[]{
      \includegraphics[width=.48\textwidth]{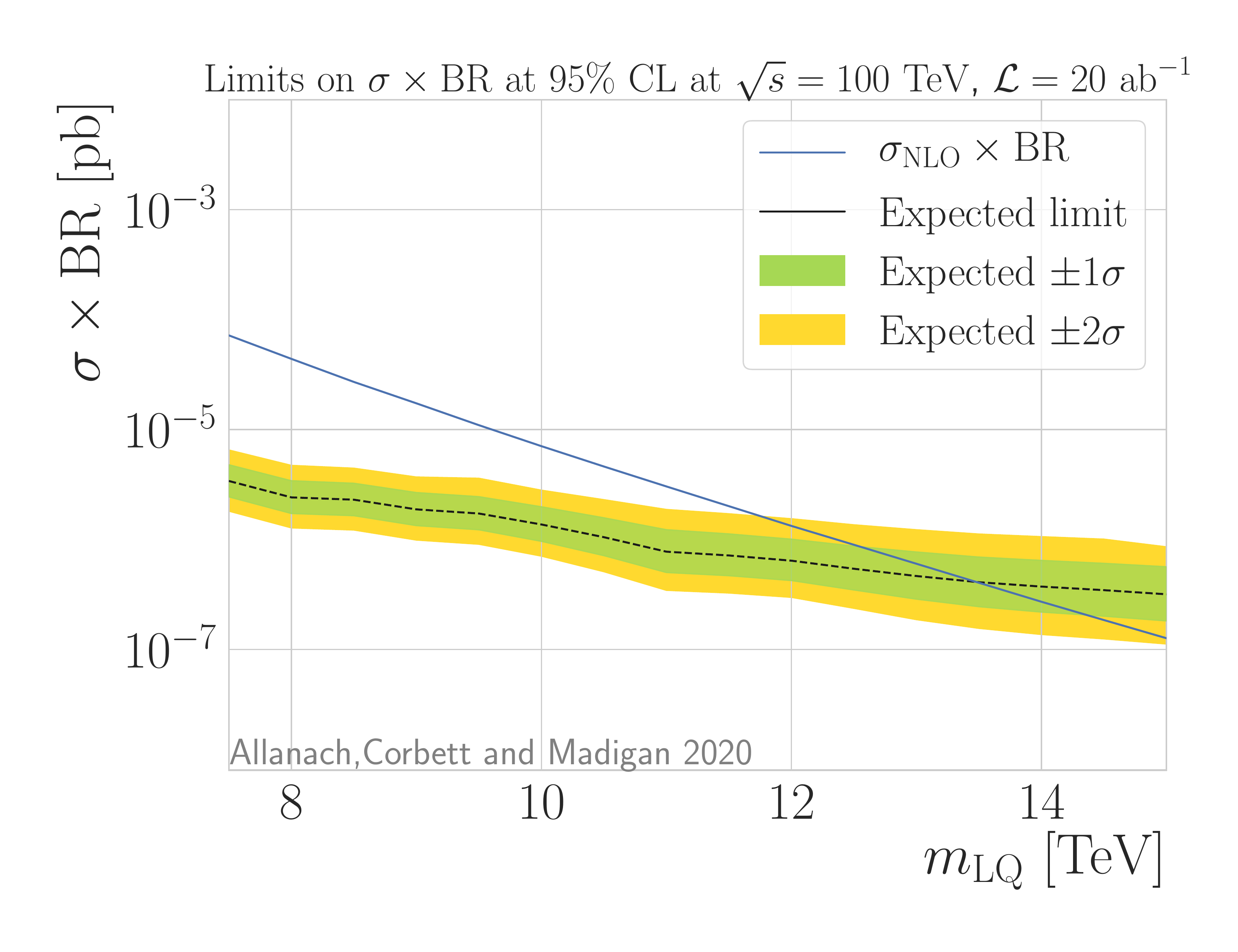}}
                \caption{Expected limits at $95 \%$ CL on $\sigma \times \text{BR}$ for the pair production of LQs and decay into a $\mu \mu j j$ final state at the HE-LHC (a) and the FCC-hh (b).  Theory curves $\sigma_{\text{NLO}} \times \text{BR}$ are calculated for narrow width LQs with couplings chosen to fit the NCBAs.}
\label{fig:limits2}
\end{figure*}

The resulting limits on $\sigma \times \text{BR}$ as a function of $m_{\text{LQ}}$ are shown in Figs. \ref{fig:limits1} and \ref{fig:limits2} for the LHC, HL-LHC, HE-LHC and FCC-hh at design integrated luminosities.  We compare our limits with theory predictions for $\sigma \times \text{BR}$, shown by the blue curves, and determine the mass to which each collider is sensitive from the point of intersection.  We see that while the sensitivity will be increased up to $m_{\text{LQ}} = 1.75$ TeV and eventually $m_{\text{LQ}} = 2.5$ TeV by the LHC Run II and HL-LHC respectively, the HE-LHC and FCC-hh have the potential to explore a much larger range of LQ parameter space, exluding masses up to $m_{\text{LQ}} = 4.8$ TeV and 13.5 TeV respectively.

To further investigate the potential of future colliders to exclude high-mass LQs, we scan over a range of integrated luminosities up to $\mathcal{L} = 3, 15$ and 20 ab$^{-1}$ for the HL-LHC, HE-LHC and FCC-hh respectively and determine the mass sensitivity at $95\%$ CL for each.  Similarly, we perform a scan over integrated luminosities and determine the discovery reach of each future collider.  These results are shown in Fig. \ref{fig:summary}.  In both plots the points correspond to the design integrated luminosities of $\mathcal{L} = 3, 15$ and 20 ab$^{-1}$.  The highest $m_{\text{LQ}}$ that can be observed with a $5 \sigma$ significance is $m_{\text{LQ}} = 9.5$ TeV: we predict that narrow width scalar LQs could be discovered at this mass assuming the FCC-hh operates at the full $\mathcal{L} = 20$ ab$^{-1}$.  Similarly, the HE-LHC and HL-LHC have the potential to observe narrow width scalar LQs of masses up to $m_{\text{LQ}} = 3.6$ TeV and $1.9$ TeV respectively.  Finally, we compute the discovery reach of the LHC Run II with $\sqrt{s} = 13$ TeV, $\mathcal{L} = 140$ fb$^{-1}$ to be $m_{\text{LQ}} = 1.2$ TeV, right on the edge of the $95 \%$ exclusion limits already found by the 13 TeV LHC as discussed in \S\ref{sec:intro}.  Table~\ref{table:nonlin} summarises the maximum $5 \sigma$ discovery reach and mass exclusion at $95 \% $ CL of each future collider.

\subsection{Wide resonances \label{sec:wide}}
If we depart from the NCBA limit, the LQs may acquire an appreciable
and non-negligible width. In this subsection, we wish to estimate how big the
effect might be on the resulting sensitivity.
Our SM
background simulations and statistical methods can be applied to
determine the approximate change in sensitivity to wider LQ resonances, where some $|y_{lq}|$ may be
large,\footnote{If any of the $|y_{lq}|$ involving quarks from the first two
  families are large, single LQ production, which is beyond the scope
  of the present paper, may also prove a profitable
search channel~\cite{Allanach:2017bta,Hiller:2018wbv}. } with one important
caveat:
we do not include interference between signal and background.  Interference can only occur between the LQ signal and DY, as all other SM background processes include neutrinos in the final state.  
Our final estimate of sensitivity will be an over-estimate because we
may expect the signal to be broadened further by signal-background interference effects. Our purpose
however, is just to see the approximate shift in sensitivity rather than to
provide a true and accurate calculation of the sensitivity itself.
We shall see that the sensitivity is not drastically changed by including
large width effects, and we expect that this qualitative conclusion
holds once signal-background interference effects have been included.

The partial decay width of a LQ into a lepton $l$ and quark $q$ is related to the mass $m_{\text{LQ}}$ and coupling $y_{lq}$ by \cite{Plehn:1997az}
\begin{align} \label{eq:LQwidth}
\Gamma = \frac{|y_{lq}|^{2} m_{\text{LQ}}}{16 \pi}.
\end{align}
Given the choice of couplings in our signal simulations as outlined in
\S\ref{sec:signal}, we have so far only considered narrow LQ
resonances satisfying $\Gamma / m_{\text{LQ}} < 0.01$.
Any narrow width LQ will still produce a wide resonance in the distribution of
$m_{\text{min}}(\mu, j)$ as shown in
Figs. \ref{fig:13TeVHLback} and \ref{fig:27TeVback}.  This is an
effect of changes in the kinematics of the final state particles due to parton
showering and detector resolution, as well as the ambiguity in defining $m_{\textrm{min}}(\mu, j)$, and determines the experimental resolution.
By fitting a normal distribution to these resonances and approximating the
resolution $\Gamma_{\text{res}}$ by twice the standard deviation, we estimate
the resolution to be $\Gamma_{\text{res}} / m_{\text{LQ}} = 0.1$.  To
investigate the effects of wide resonances we then simulate LQ events with
decay width $\Gamma \geq \Gamma_{\text{res}}$.  We do this by switching on the
same couplings $(Y_{de})_{22} = (Y_{de})_{32} \neq 0$ as before, determining
their values from Eq.~\ref{eq:LQwidth} for $\Gamma = 0.1, 0.2$ and $0.5$.

\begin{figure*}[htp]
   \subfloat[]{
      \includegraphics[width=.48\textwidth]{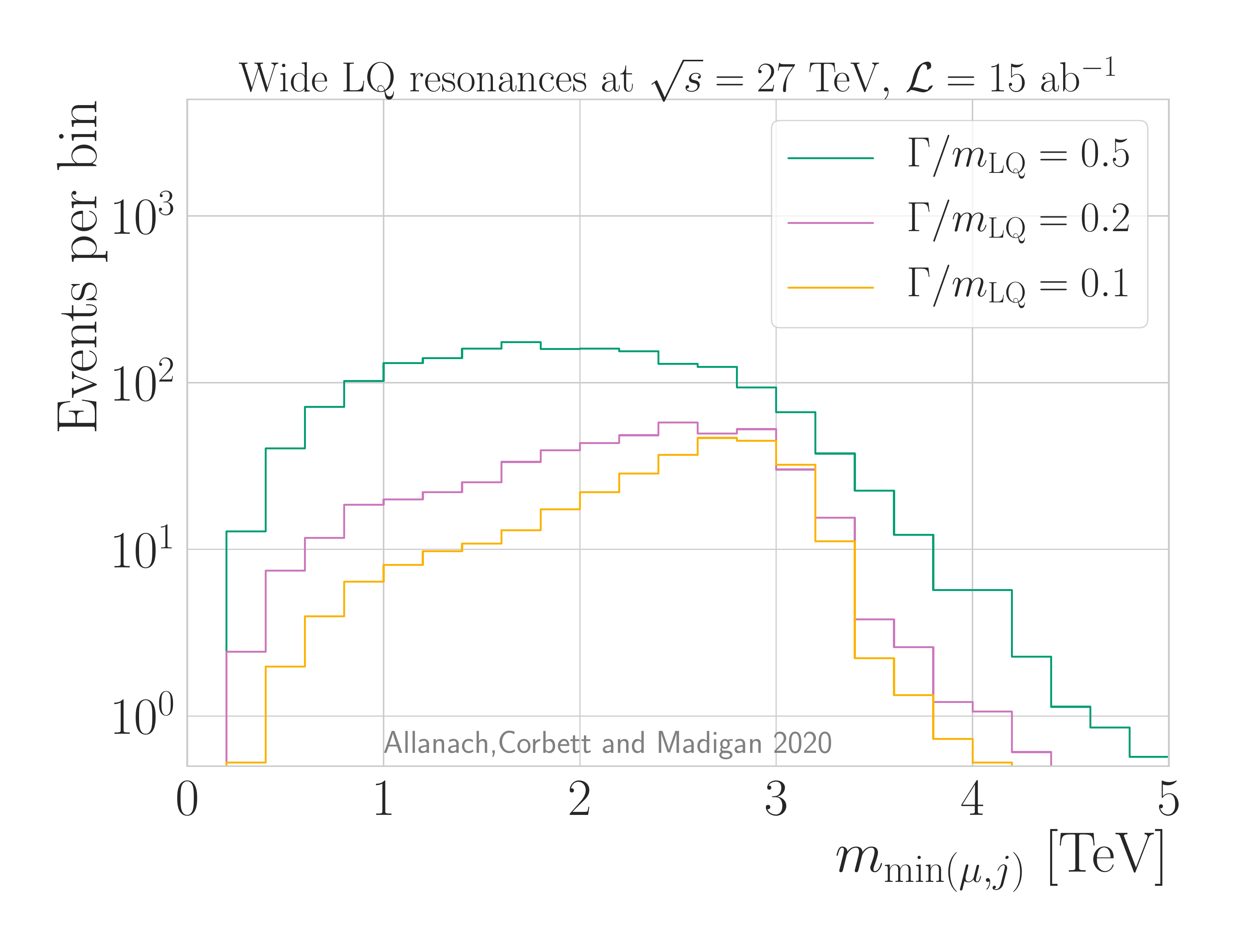}}
   \subfloat[]{
      \includegraphics[width=.48\textwidth]{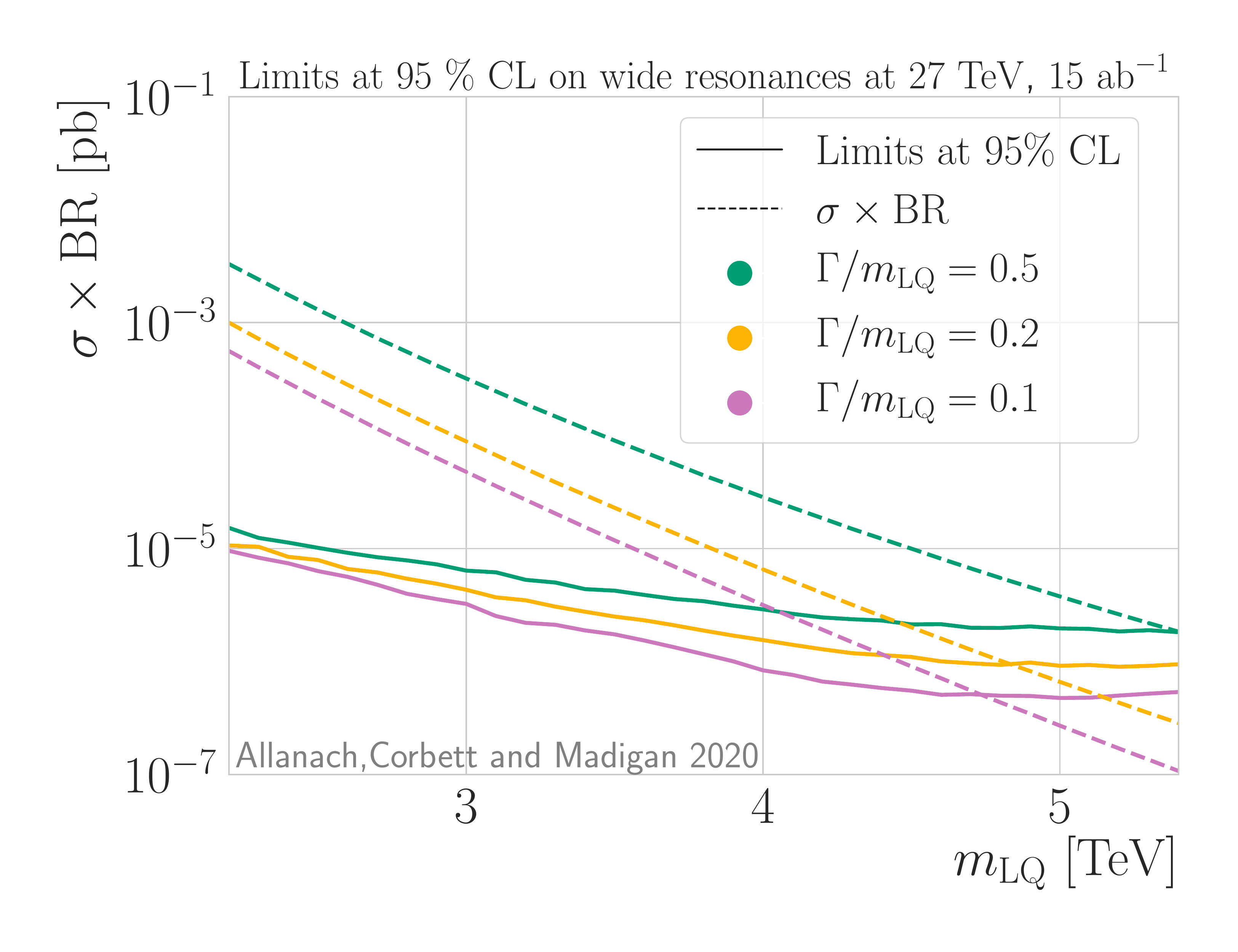}}
  \caption{\label{fig:wideLQ1} (a): Comparison of the predicted $m_{\text{min}}(\mu, j)$ distribution of LQ signal events for a LQ with large decay width $\Gamma$.  (b): Expected limits at $95 \%$ CL on $\sigma \times \text{BR}$ for the pair production of wide LQs decaying into a $\mu \mu j j$ final state at $\sqrt{s} = 27$ TeV, $\mathcal{L} = 15$ ab$^{-1}$ (b).}
\end{figure*}

\begin{figure*}[htp]
   \subfloat[]{
      \includegraphics[width=.48\textwidth]{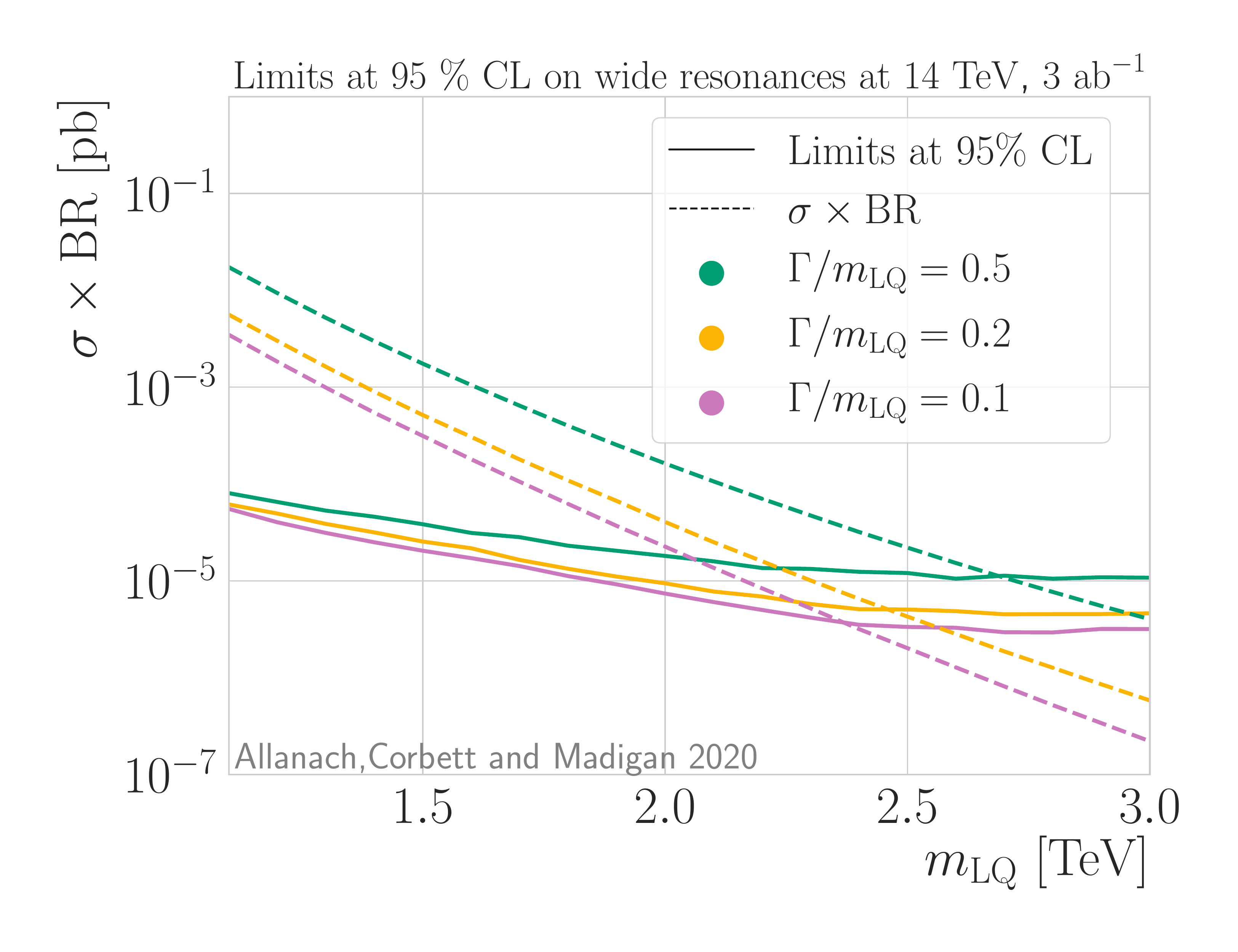}}
   \subfloat[]{
      \includegraphics[width=.48\textwidth]{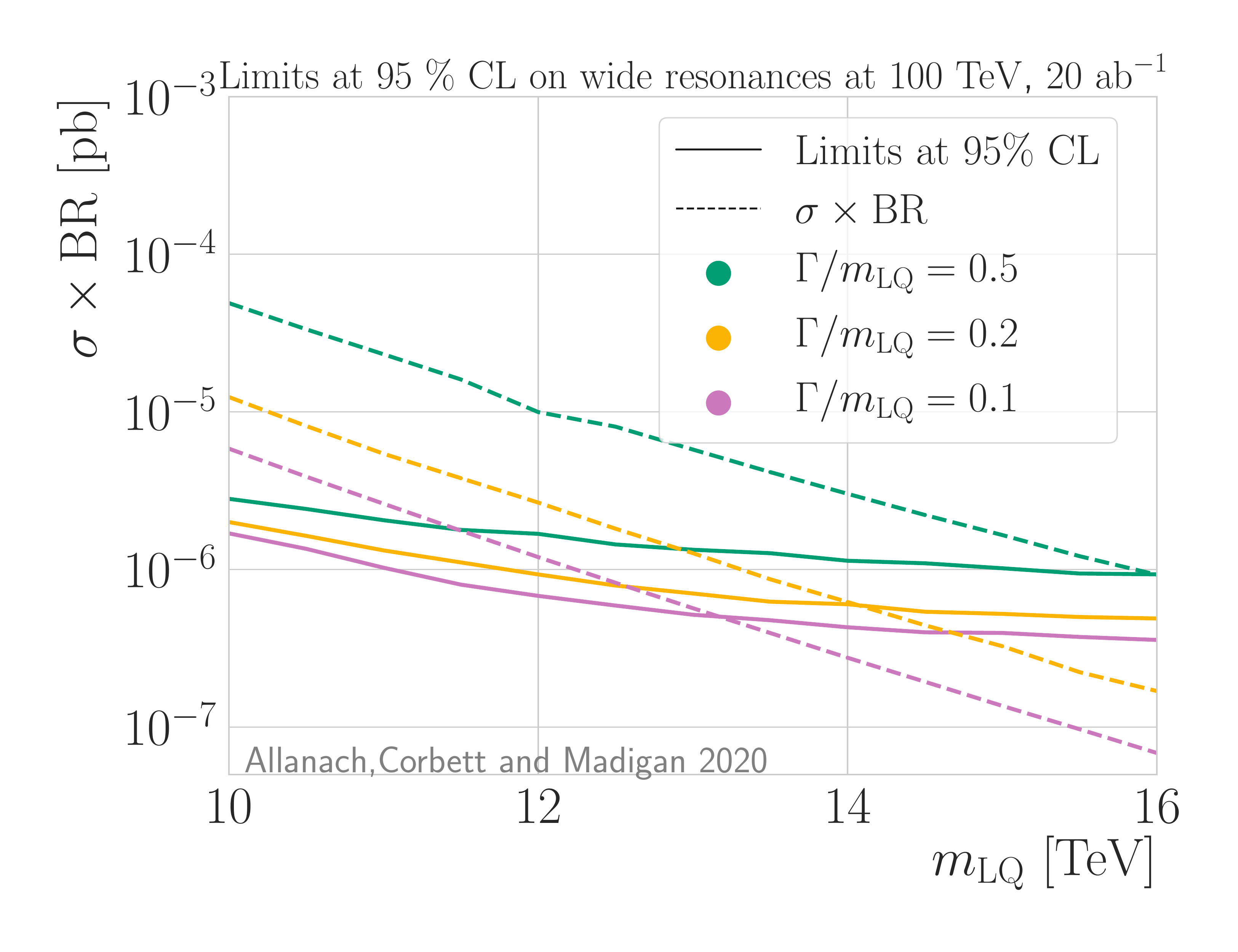}}
  \caption{\label{fig:wideLQ2} Expected limits at $95 \%$ CL on $\sigma \times \text{BR}$ for the pair production of LQs with large decay width $\Gamma$, decaying into a $\mu \mu j j$ final state at $\sqrt{s} = 14$ TeV, $\mathcal{L} = 3$ ab$^{-1}$ (a) and $\sqrt{s} = 100$ TeV, $\mathcal{L} = 20$ ab$^{-1}$ (b).}
\end{figure*}

Fig. \ref{fig:wideLQ1} (a) compares our simulations of the\\ $m_{\text{min}}(\mu, j)$ distributions of large width LQs at $m_{\text{LQ}} = 3.2$ TeV for $\sqrt{s} = 27$ TeV, $\mathcal{L} = 15$ ab$^{-1}$ and $\Gamma/m_{\text{LQ}} = 0.1,0.2$ and $0.5$.  Fig. \ref{fig:wideLQ1} (b)  shows the corresponding expected limits on $\sigma \times \text{BR}$ at $95 \%$ CL for the HE-LHC at 27 TeV, 15 ab$^{-1}$.  To provide a sample estimate of sensitivity we compare our limits to the values of $\sigma_{\text{LO}} \times \text{BR}$ calculated for a wide LQ signal with nonzero couplings $(Y_{de})_{22}$ and $(Y_{de})_{32}$ as outlined above.

Fig. \ref{fig:wideLQ1}  shows that the distribution of signal events spreads out in $m_{\text{min}}(\mu, j)$ with increasing $\Gamma$.  We expect that the sensitivity to LQs is decreased as a result of the signal events spreading out in this way rather than being peaked around a few bins.  This effect is seen in the increase in the upper limits on $\sigma \times \text{BR}$ with increasing $\Gamma$ in Fig. \ref{fig:wideLQ1} (b).  We can see from the intersection of the $\Gamma / m_{\text{LQ}} = 0.1$ theory curve (purple, dashed) with each set of expected limits (solid curves) that if the theory predictions for $\sigma \times \text{BR}$ were independent of LQ couplings $(Y_{de})_{22}$ and $(Y_{de})_{32}$, an increase in LQ width from $\Gamma / m_{\text{LQ}} = 0.1$ to $\Gamma / m_{\text{LQ}} = 0.5$ would result in a loss of sensitivity from approximately $m_{\text{LQ}} = 4.8$ TeV to $m_{\text{LQ}} = 4$ TeV.  However, this effect is mitigated by the fact that at such large couplings, pair production is no longer dominated by gluon-gluon interactions.  Instead, pair production via quark-lepton interactions has a significant contribution to the total cross section.  As a result, the theory prediction for $\sigma \times \text{BR}$ depends strongly on the choice of couplings $(Y_{de})_{22}$ and $(Y_{de})_{32}$.  This can be seen by the overall increase in the number of signal events with $\Gamma$ in Fig. \ref{fig:wideLQ1} (a), and by the large increase in the values of $\sigma \times \text{BR}$ with $\Gamma$ in Fig. \ref{fig:wideLQ1} (b).  Overall this leads to an increase in sensitivity to LQs with increasing $\Gamma$.  A similar effect is seen in our predictions for wide LQs at the HL-LHC and FCC-hh, as shown in Fig. \ref{fig:wideLQ2}.
Were we to include signal-background interference effects in the calculation, the sensitivity
would be degraded. This leads us to conclude that
the overall change in sensitivity from the larger
widths is not dramatic and would remain small were we to include signal-background
interference effects.

\section{Conclusions}
\label{sec:conc}

\begin{figure*}[htp]
   \subfloat[]{
      \includegraphics[width=.48\textwidth]{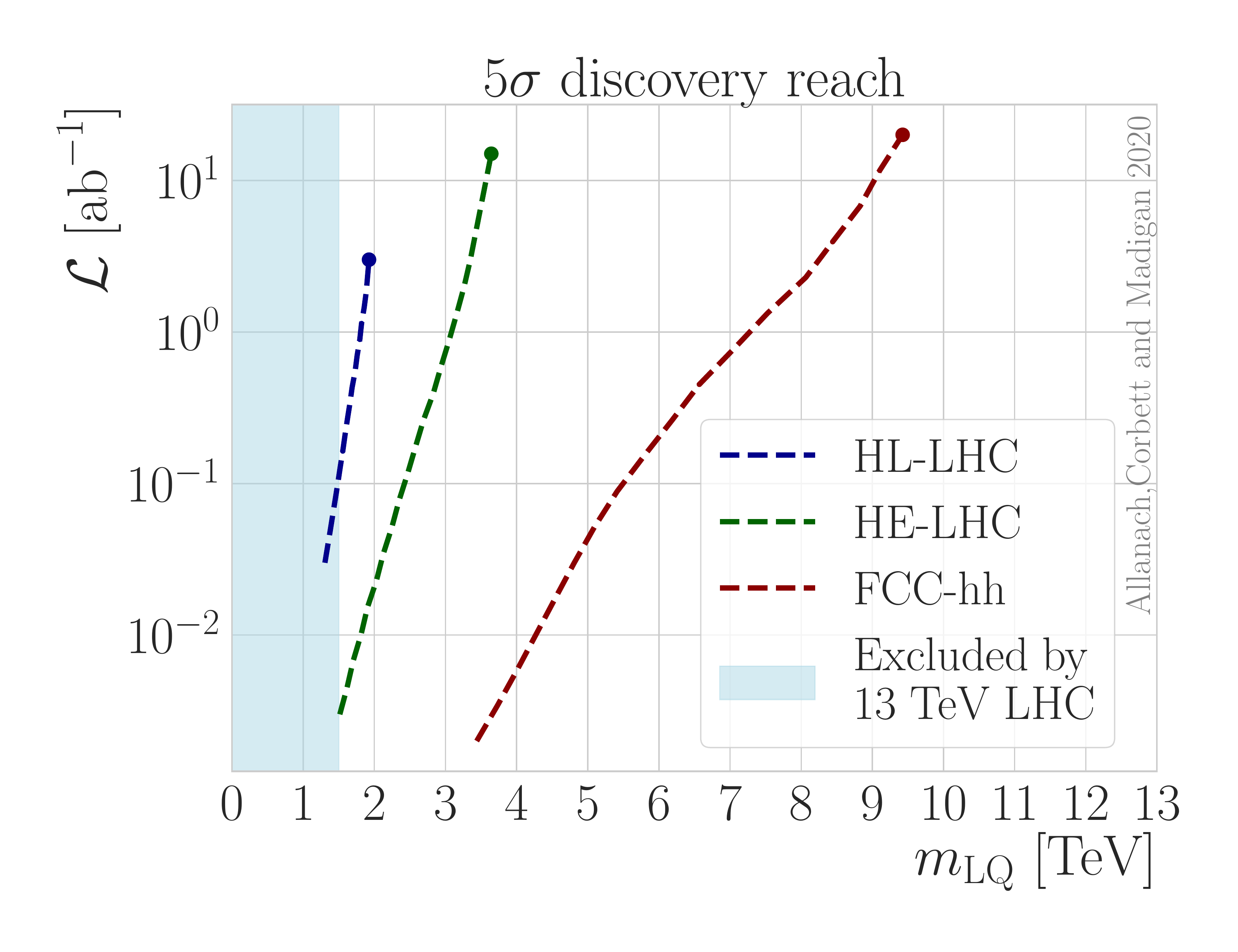}}
   \subfloat[]{
      \includegraphics[width=.48\textwidth]{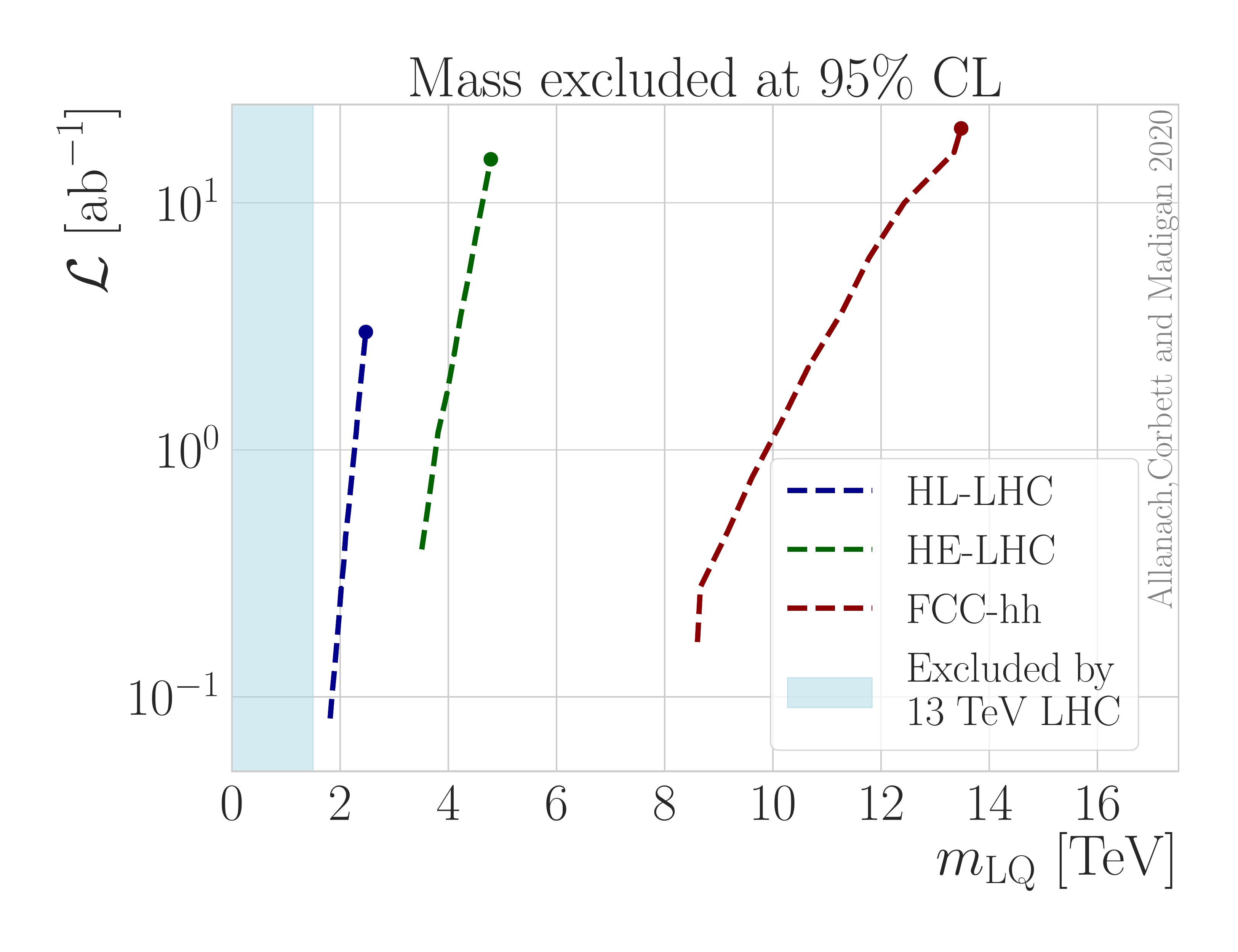}}
	\caption{Predicted $5 \sigma$ discovery reach (a) and mass exclusion at $95 \%$ CL (b) of the HL-LHC, HE-LHC and FCC-hh.  Points correspond to the design integrated luminosities of each future collider of $\mathcal{L} = 3,15,20$ ab$^{-1}$ respectively. \label{fig:summary}}
\end{figure*}

We have estimated the exclusion and discovery sensitivities of future hadron
colliders to LQ pair production for the case that each LQ decays to a muon and a
jet. Such a decay channel is motivated in part by the LQ solution to the
NCBAs. It is also motivated by the fact (regardless of
the NCBAs)
that muons are
empirically robust objects, which are good for tagging and beating down
irreducible backgrounds.
By concentrating on LQ pair production (rather than single LQ production, for
example) we cover a large volume of model parameter space where LQs, being
perturbatively coupled, are narrow
and the pair production cross-section varies only with
the LQ mass $m_{\text LQ}$.
For such LQs, their production is dominated by production from glue-glue
interactions, their interactions with initial state quarks being negligible.
This is typically true for LQs that have a
coupling-mass relation consistent with the NCBAs, but we emphasise that our
sensitivities extend beyond this coupling-mass relation more generally, as
discussed below.

The previous estimate of the exclusion sensitivity in Ref.~\cite{Allanach:2017bta}
extrapolated LHC search limits using two highly dubious approximations. The first being that experimental
efficiency and acceptance would not change with centre of mass energy, and the second that LQs are produced exactly at threshold. With respect to the first point,
at large $m_{\rm LQ}$ and at high energies (particularly at FCC-hh), the decay products
from LQs will be highly boosted resulting in muons
collinear to the jets resulting in more muons failing isolation
criteria. The muon momentum resolution is also likely to be very
poor at higher energies, since such hard muons will only be bent to a limited extent by the
magnets. This also affects signal efficiency due to peak broadening. Secondly, the assumption LQs are produced exactly at threshold is likely to introduce large uncertainties\footnote{Ref.~\cite{Allanach:2017bta}
  also considered narrow $Z^\prime$ production, for which the second
  approximation should be more accurate. The methodology employed therein
  therefore suited the $Z^\prime$
case much better.}.

\begin{table}[h]
\centering
\begin{tabular}{|c |c |c |c |c |}
\hline
  &   &    & $5 \sigma$ disc. & Mass excl.\\ [0.5ex]
	Collider	& $\sqrt{s}$  & $\mathcal{L}$  & reach & at $95 \%$ CL\\ [0.5ex]
	 &[TeV] & $[ab^{-1}]$ & [TeV] & [TeV]\\ [0.5ex]
\hline \hline
	LHC  & 13 & 0.14 & 1.2$^\ast$ & 1.8\\
\hline
 HL-LHC & 14 & 3 & 1.9 &2.5 \\
 \hline
HE-LHC  & 27 & 15 & 3.6 & 4.8 \\
 \hline
FCC-hh  & 100 & 20 & 9.5 & 13.5  \\
\hline
\end{tabular}
\caption{Summary of the expected $5\sigma$ discovery sensitivity and expected 95$\%$ CL
  exclusion
 sensitivity to $S_3$
for future hadron colliders, from LQ pair production.  ${}^\ast$The predicted discovery
  reach of the 13 TeV LHC at 140 fb$^{-1}$ of $m_{\rm{LQ}} = 1.2$ TeV is
  currently on the edge of exclusion at $95 \%$ CL (see
  Fig.~\protect\ref{fig:13TeVlimits}).  }
\label{table:nonlin}
\end{table}
We rectify these two bad
approximations in our\\ present paper by performing a fast simulation of the signal and
detector response.
We summarise our expected discovery and exclusion sensitivities in
Table~\ref{table:nonlin}.
Ref.~\cite{Allanach:2017bta} estimated
that the HL-LHC could exclude 2.2 TeV at 95$\%$ CL, to be compared with 1.8
TeV. The HE-LHC was
estimated to cover up to 4.1 TeV, but this was for a higher centre of mass
energy (33 TeV) and a different luminosity (15 ab$^{-1}$), precluding a direct
comparison. The FCC-hh exclusion sensitivity was calculated at an integrated
luminosity of 10 ab$^{-1}$ to be 12.0 TeV,
to be compared with 12.5 TeV from our estimate (see Fig.~\ref{fig:summary}).
It is somewhat surprising that the comparable estimates are so similar, since
as we have argued, the old ones were based on self-admitted bad approximations.
The results in Table~\ref{table:nonlin} are on a much firmer footing.
This is the first time that the
$5 \sigma$ $S_3$ discovery sensitivities for future colliders have appeared.
It  is also the first time that $S_3$ sensitivity estimates for varying
luminosities
have been calculated as in Fig.~\ref{fig:summary}.

The sensitivities phrased in terms of LQ mass have a dependence on the
LQ-fermion couplings
assumed in the model, since these may affect the BR of the muon-jet decay
rate. However, all limits in the narrow LQ limit on $\sigma \times BR$ also apply
to models with different (but still small) LQ couplings to fermions.
Only
when one or more of the LQ couplings approaches the non-perturbative r\'{e}gime does the LQ
width become comparable to the experimental resolution, potentially affecting sensitivity.
To cover this case,
we considered a wider LQ: see
\S\ref{sec:wide}. Of and by itself, the width does not change the sensitivity much. Increasing the width
divided by mass of the LQ from 0.1 to 0.5 but keeping the
cross-section times branching ratio constant only results in a 10$\%$ degradation
or so in FCC-hh mass reach, as the right-hand panel of Fig.~\ref{fig:wideLQ1}
shows.

We hope that the results of our study will be useful for the
current European Strategy in Particle Physics ~\cite{Heinemann:2691414} and
provide a part of the physics case for future hadron
colliders~\cite{CidVidal:2018eel,Abada:2019ono,Abada:2019lih,Benedikt:2018csr}.

\iffalse
\section{Introduction}
\label{intro}
Your text comes here. Separate text sections with
\section{Section title}
\label{sec:1}
Text with citations \cite{RefB} and \cite{RefJ}.
\subsection{Subsection title}
\label{sec:2}
as required. Don't forget to give each section
and subsection a unique label (see Sect.~\ref{sec:1}).
\paragraph{Paragraph headings} Use paragraph headings as needed.
\begin{align}
a^2+b^2=c^2
\end{align}

\begin{figure}
  \includegraphics{example.eps}
\caption{Please write your figure caption here}
\label{fig:1}       % Give a unique label
\end{figure}
\begin{figure*}
  \includegraphics[width=0.75\textwidth]{example.eps}
\caption{Please write your figure caption here}
\label{fig:2}       % Give a unique label
\end{figure*}
\begin{table}
\caption{Please write your table caption here}
\label{tab:1}       % Give a unique label
\begin{tabular}{lll}
\hline\noalign{\smallskip}
first & second & third  \\
\noalign{\smallskip}\hline\noalign{\smallskip}
number & number & number \\
number & number & number \\
\noalign{\smallskip}\hline
\end{tabular}
\end{table}
\fi
\begin{acknowledgements}
We thank the ATLAS exotics group and
other members of the\/ {\em Cambridge Pheno Working Group} for
helpful advice and comments, especially T You for constructive criticism of
the draft.
This work has been partially supported by STFC HEP consolidated grant
ST/P000681/1. TC acknowledges support from the Villum Fonden and the Danish National Research Foundation (DNRF91) through the Discovery center.  MM acknowledges support from the Schiff Foundation.
\end{acknowledgements}
\bibliographystyle{spphys}
\bibliography{LQ}
\end{document}